\documentclass{siamltex}
\usepackage{euscript,amsmath,amssymb,amsfonts,graphicx,bm}
  \usepackage{paralist}
  \usepackage{epstopdf}
  \usepackage{graphics,color} 
 \usepackage[colorlinks=true]{hyperref}
 \hypersetup{urlcolor=blue, citecolor=red}
\usepackage{cite}

\newcommand{\e}{{\rm e}}
\renewcommand{\d}{{\rm d}}

\newcommand{\x}{{\mathbf x}}

\newcommand{\D}{\displaystyle}

\newcommand{\sign}{{\rm sign}}
\newcommand{\g}{{\mathbf g}}
\newcommand{\W}{{\mathbf W}}

\definecolor{darkgreen}{rgb}{0,0.4,0}

\title{Stochastic models of evidence accumulation in changing environments}

\author{Alan Veliz-Cuba\thanks{Department of Mathematics, University of Dayton, Dayton, OH 45469 USA ({\tt avelizcuba1@udayton.edu})} \and Zachary P. Kilpatrick$^\Diamond$\thanks{Department of Mathematics, University of Houston, Houston, TX 77204 USA ({\tt zpkilpat@math.uh.edu}). $\Diamond$ ZPK and KJ contributed equally to this work.} \and Kre\v{s}imir Josi\'{c}$^\Diamond$\footnotemark[2]  \thanks{Department of Biology and Biochemistry, University of Houston, Houston, TX 77204 USA ({\tt josic@math.uh.edu})}. $\Diamond$ ZPK and KJ contributed equally to this work.}

\date{\today}

\begin{document}

\maketitle

\begin{abstract} Organisms and ecological groups accumulate evidence to make decisions. Classic experiments and theoretical studies have explored this process when the correct choice is fixed during each trial. However, we live in a constantly changing world. What effect does such impermanence have on classical results about decision making?  To address this question we use sequential analysis to derive a tractable model of evidence accumulation when the correct option changes in time. Our analysis shows that ideal observers discount prior evidence at a rate determined by the volatility of the environment, and the dynamics of evidence accumulation is governed  by the information gained over an average environmental epoch. A plausible neural implementation of an optimal observer in a changing environment shows that, in contrast to previous models,  neural populations representing alternate choices are coupled through excitation. Our work builds a bridge between  statistical decision making in volatile environments and stochastic nonlinear dynamics.

\end{abstract}

\section{Introduction} 

To navigate a constantly changing world, we intuitively use the most recent and pertinent information. For instance, when planning a route between home and work we use recent reports of accidents and weather. We discount older information, as our environment is in constant flux: The clouds threatening rain last night may have dissipated, and an accident reported an hour ago has likely been cleared.   The optimal strategy is therefore  to  weight recent evidence more strongly.  
 
 How to make decisions in face of uncertainty and impermanence is a question that recurs in fields ranging from economics to  ecology and neuroscience.
Mammals \cite{busemeyer93,bogacz06,gold07,brunton13}, insects~\cite{seeley91,couzin05}, single cells \cite{balazsi11}, and
animal collectives~\cite{Marshall09} gather evidence to make decisions.  However information about the state of the world is typically incomplete and perception is noisy.  
Therefore, animals make choices based on uncertain evidence.  
The case of an observer deciding between two alternatives based on a series of noisy measurements has been studied extensively when the environment is static~\cite{neyman33,wald48,gold02,bogacz06}. In this case humans~\cite{ratcliff04}, and other mammals~\cite{gold07,brunton13} 
can accumulate incoming evidence near optimally to reach a decision. 

Stochastic accumulator models provide a plausible neural implementation of decision making between two or more alternatives~\cite{usher01,beck08}. 
These models are analytically tractable~\cite{bogacz06}, and can implement optimal decision strategies~\cite{bogacz07}. Remarkably, there is also a parallel between these  models and experimentally observed neural activity. Recordings in animals during a decision task suggest that neural activity reflects the 
weight of evidence for one of the choices~\cite{gold07}.

A key assumption in many models is that the correct choice is fixed in time, \emph{i.e.} decisions are made in 
a static environment. This assumption may hold in the laboratory, but natural environments are seldom static~\cite{domenici97,passino08}. Recent experimental evidence suggests that human observers integrate
noisy measurements near optimally even 
when the state of the environment changes.  For instance, when observers need to decide between two options and
the corresponding reward changes in a history-dependent manner, human behavior approximates that of a Bayes optimal observer~\cite{behrens07}. An important feature of evidence accumulation in volatile environments is an increase in learning rate when recent observations do not support a current estimate~\cite{nassar12}. Both behavioral and fMRI data show that human subjects employ this strategy when they must predict the position of a stochastically moving target~\cite{mcguire14}. Experimental work thus suggests that humans adjust evidence valuation to account for environmental variability.

However, the dynamics of decision making in changing environments has not been fully investigated.   To address this question we extend optimal stochastic accumulator models to a changing environment.
These extensions are amenable to analysis, and reveal that an optimal observer discounts old 
information at a rate adapted to the frequency of
environmental changes.  As a result, the certainty that can be attained about any of the choices is limited. Our approach frames the decision making process in terms of a first passage problem for a doubly stochastic nonlinear model that can be examined using techniques of nonlinear dynamics. Extending previous work, we also identify accurate piecewise linear approximations to the nonlinear model. This model also suggests a biophysical neural implementation for evidence integrators consisting of neural populations whose activity represents the evidence in favor of a particular choice.  When the environment is not static, optimal evidence discounting can be performed exactly by populations  coupled through excitation. We also show that the computation can be well approximated by appropriately tuned classical linear population models~\cite{mcmillen06,bogacz07,usher04,smith04}. 


\section{Optimal decisions in a static environment} \label{S:static}

We develop our model in a way that parallels the case of a  static environment with two possible states. 
We therefore start with  the derivation of the recursive equation for the log-likelihood ratio of the two states, and the approximating stochastic differential equation (SDE), when the underlying state is fixed in time.

To make a decision, an optimal observer integrates a stream of measurements to infer the present environmental state.  In the static case, this
can be done using sequential analysis~\cite{wald48,busemeyer93}:  An observer makes a stream of independent, noisy measurements, $\xi_{1:n} = (\xi_1,\xi_2,...,\xi_n),$ at equally spaced times, $t_{1:n} = (t_1,t_2,...,t_n)$. The probability of each measurement, $f_{+}(\xi_n):={\rm Pr}(\xi_n | H_+)$, and $f_-(\xi_n):={\rm Pr}(\xi_n | H_-),$ depends on the environmental state. Combined with the prior probability, ${\rm Pr}(H_{\pm}),$ of the states, this gives the ratio of probabilities, 
\begin{equation*}
R_n = \frac{{\rm Pr}(H_+ | \xi_{1:n})}{ {\rm Pr}(H_- | \xi_{1:n})} =  \frac{f_+(\xi_1)f_+(\xi_2) \cdots f_+(\xi_n)}{f_-(\xi_1) f_-(\xi_2) \cdots f_-(\xi_n)} \frac{{\rm Pr}(H_+)}{{\rm Pr}(H_-)}, 
\end{equation*}
which can also be written recursively \cite{wald48}:
\begin{equation}
R_n = \left( \frac{f_+(\xi_n)}{f_-(\xi_n)} \right) \cdot R_{n-1},   \label{recLR}
\end{equation}
where  $R_0= {\rm Pr}(H_+)/{\rm Pr}(H_-)$ can describe an observer's prior belief about the probability of the
two choices.

With a fixed number of observations, the ratio in Eq.~\eqref{recLR} can be used to make a choice that
minimizes the total error rate~\cite{neyman33}, or maximizes reward~\cite{gold02}. 
Eq.~\eqref{recLR}  gives a recursive relation for the log-likelihood ratio, $y_n = \ln R_n,$
\begin{align}
y_n = y_{n-1} + \ln \frac{f_+(\xi_n)}{f_-(\xi_n)}. \label{recIn}
\end{align}
When the time between observations, $\Delta t = t_j - t_{j-1},$ is small, we can use the Functional Central Limit Theorem~(p.~357~ in~\cite{Billingsley_book}) to approximate this stochastic process by the stochastic differential equation (SDE)~\cite{ratcliff78,bogacz06}, 
\begin{align}\label{eqn:sde_DD}
dy = g_{\pm} dt + \rho_{\pm} dW_t, 
\end{align}
where  $W_t$ is a Wiener process, and the constants $g_{\pm}= \frac{1}{\Delta t} {\rm E}_{\xi} [\ln \frac{f_+(\xi)}{f_-(\xi)} | H_\pm]$  and  $ \rho^2_{\pm}= \frac{1}{\Delta t}\text{Var}_{\xi} [\ln \frac{f_+(\xi)}{f_-(\xi)}| H_\pm]$  depend on the environmental state.
Below we approximate other discrete time process, such as that given by Eq.~\eqref{recIn}, with 
SDEs. Details of these derivations are provided in the Appendix.

In state $H_+$ we have $
g_+ \Delta t =   \int_{-\infty}^{\infty} f_+ ( \xi) \ln \frac{f_+(\xi)}{f_-(\xi)} \d \xi
$. The drift between two observations thus equals the Kullback--Leibler divergence between $f_+$ and $f_-,$ 
\emph{i.e.} the strength of the observed evidence from a measurement in favor of $H_+$~\cite{cover12}.  An equivalent interpretation
holds for $g_-$. 
Hence $g_+$ and $g_-$ are the rates at which an optimal observer accumulates information.  We will use this observation
to interpret the parameters of the model in a changing environment.

\section{Two alternatives in a changing environment} 

\begin{figure}
\centering
\includegraphics[width=3.3in]{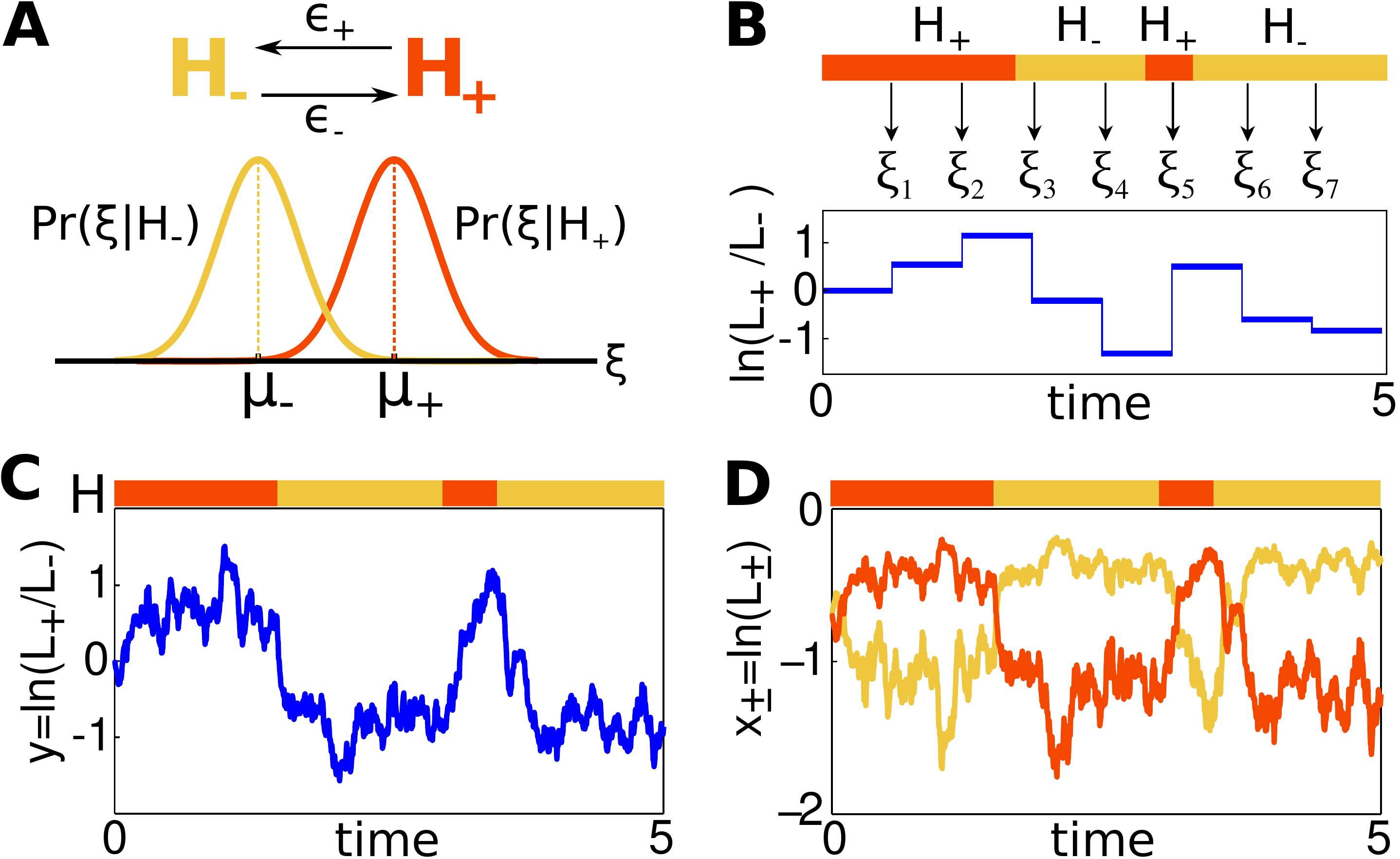}
\vspace{-.1cm}
 \caption{Evidence accumulation in a changing environment. 
 (\textbf{A})
The environmental state transitions from state $H_+$ to $H_-$ and back with rates $\epsilon_+$ and $\epsilon_-$, respectively. Observations 
follow state dependent probabilities, $f_\pm(\xi)=\text{Pr}(\xi|H_\pm)$.
 (\textbf{B})
The distributions of the measurements, $\xi_n,$ change with the environmental state. Each individual observation
changes the  log-likelihood ratio, $\ln (L_{n,+}/L_{n,-})$.  A single realization is shown.
 (\textbf{C,D})
The evolution of the continuous approximation of the log-likelihood ratio, $y(t)$, (panel \textbf{C}) and the log probabilities 
$x_\pm(t)$ (panel \textbf{D}). At time $t$, evidence favors the environmental state  $H_+$ if $y(t)>0,$ or, equivalently, if $x_+(t)>x_-(t)$.
\vspace{-0.3cm}
 }
\label{fig:fig1}
\end{figure}

We use the same assumptions to derive a recursive equation for the log-likelihood ratio between two alternatives in a changing environment. The state of the environment, $H(t),$ is $H_+$ or  $H_-$, but can now change in time (See Fig.~\ref{fig:fig1}A,B). When the environment is in one of these two possible states, the statistics of the observations are fixed. Observation statistics are  therefore piecewise stationary in time. An observer infers the present state from a sequence of observations, $\xi_{1:n}$, made at equally spaced times, $t_{1:n}$ with $\Delta t = t_j - t_{j-1}$ and characterized by probabilities $f_{\pm} (\xi_n) : = {\rm Pr}(\xi_n| H_{\pm})$. The state of the environment 
changes according to a telegraph process (\emph{e.g.}, p.~77~in~\cite{gardiner04}), and the probability of a change between two observations is $\epsilon_{\pm} \Delta t  := {\rm Pr}(H(t_n) = H_{\mp} | H(t_{n-1}) = H_{\pm} )$. We assume that $\epsilon_+$ and $\epsilon_-$ are known to the observer.

The probabilities, $L_{n,\pm} =\text{Pr}(H( t_n) = H_{\pm}|\xi_{1:n}),$ then satisfy (See Appendix \ref{llr2app}):
\begin{align} \label{eqn:general_disc_likelihood}
L_{n,\pm} \propto f_{\pm}(\xi_n)
\left( (1-\Delta t \epsilon_{\pm}) L_{n-1,\pm} + \Delta t \epsilon_{\mp} L_{n-1,\mp} \right),
\end{align}
with proportionality constant ${\rm Pr}(\xi_{1:n-1})/{\rm Pr}(\xi_{1:n})$.
As in the static case, the  ratio of the probabilities of the two environmental states at time $t_n$, can be determined recursively (See Appendix \ref{llr2app}), and
equals
\begin{equation} \label{eqn:general_disc}
R_n  = \frac{L_{n,+}}{L_{n,-}} = \frac{f_+(\xi_n)}{f_-(\xi_n)} 
\frac{(1-\Delta t \epsilon_+) R_{n-1} + \Delta t \epsilon_{-} }
{ \Delta t \epsilon_{+} R_{n-1} + 1- \Delta t \epsilon_{-}  }.
\end{equation}
In this expression, the ratio of probabilities at the time of the previous observations, $R_{n-1},$ is  discounted in a way that
depends on the frequency of environmental changes, $\epsilon_{\pm}$. This equation, and the continuum limits we discuss below, have been derived previously \cite{deneve08,zhao10}, but their dynamics were not analyzed.

\begin{figure}
\centering
\includegraphics[width=3.3in]{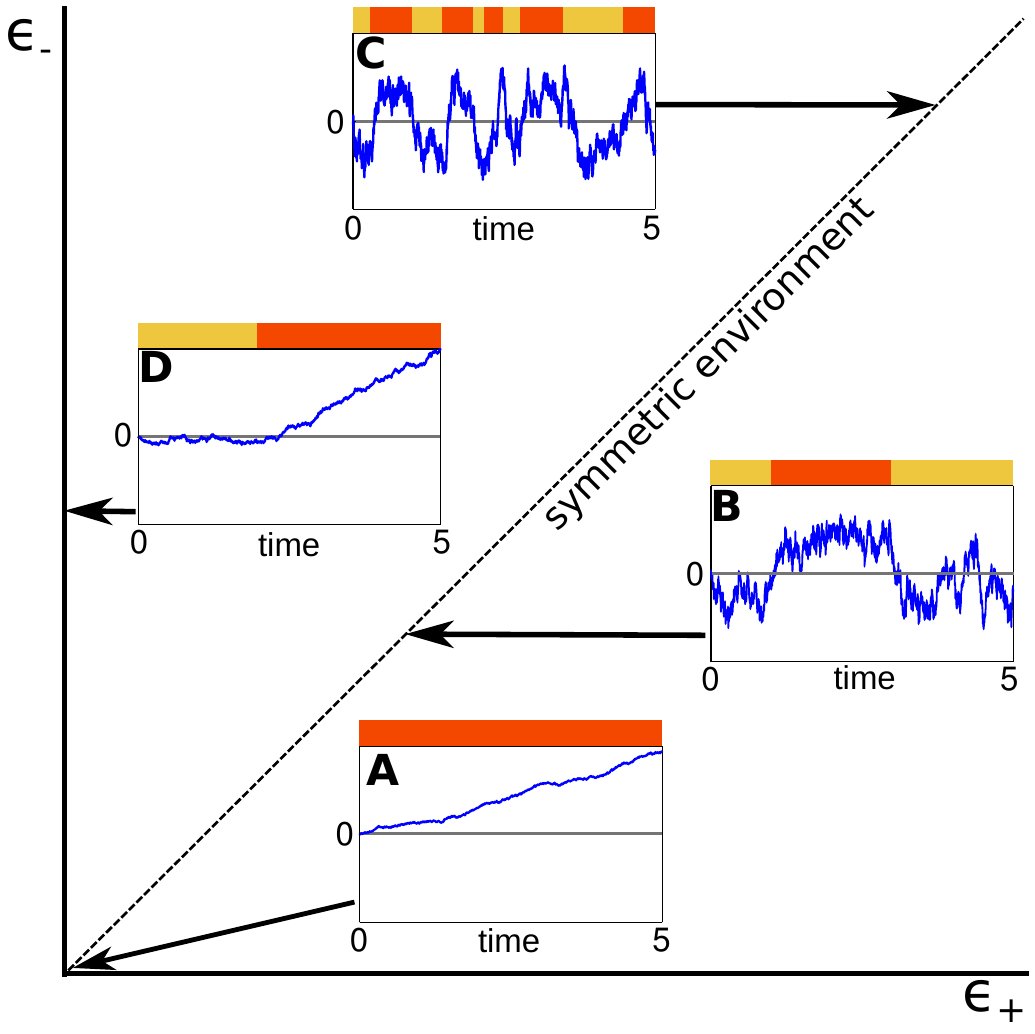}
 \caption{In a dynamic environment, the dynamics of the log-likelihood ratio, $y,$ depends on the rates of switching between states. 
 ({\bf A}) When $\epsilon_{\pm} = 0$, the environment is static, and the model reduces to the one derived in Section~\ref{S:static}. ({\bf B}) When then environment changes slowly, $|\epsilon_{\pm}| \ll 1$, the log-likelihood ratio, $y,$ can saturate. ({\bf C}) In
 a rapidly changing environment, $y$ tends  not to equilibrate. ({\bf D}) When $\epsilon_+ = 0$ and $\epsilon_- >0$, the task becomes a change detection problem. }
\label{fig:fig2}
\end{figure}

Eq.~\eqref{eqn:general_disc} describes a variety of cases of evidence accumulation studied previously (See Fig.~\ref{fig:fig2}): If the environment is fixed ($\epsilon_{\pm}=0$), we recover Eq.~\eqref{recLR}. If the environment starts in state $H_-$, changes to $H_+$, but cannot change back ($\epsilon_->0, \epsilon_+=0$),  we obtain
\begin{equation*} 
R_n  = \frac{f_+(\xi_n)}{f_-(\xi_n)} 
\frac{R_{n-1} + \Delta t \epsilon_{-} }
{1- \Delta t \epsilon_{-}  },
\end{equation*}
a  model used in change point detection~\cite{yu06,adams07,shiryaev07}.  

We can again approximate evolution of $y_n = \ln R_n$, \emph{i.e.} the stochastic process 
describing the evolution of the log of the likelihood ratio in Eq.~\eqref{eqn:general_disc}, by an SDE:
\begin{align}
\d y & = [g(t) + \underbrace{\epsilon_{-}(e^{-y}+1)-\epsilon_{+}(e^y+1)}_{\text{nonlinearity}}] \d t + \rho(t) \d W_t  , \nonumber \\[-3.5ex] \ \label{eqn:sde_two}
\end{align}
where 
$
g (t)=  \frac{1}{\Delta t}  
{\rm E}_{\xi} \left[ \ln \frac{f_+(\xi)}{f_-(\xi)}  \bigg| H(t) \right] \!\!,
$
and 
$
\rho^2(t)=  \frac{1}{\Delta t} 
\text{Var}_{\xi} \left[ \ln \frac{f_+(\xi)}{f_-(\xi)}  \bigg| H(t) \right]
$. Note that the drift and variance 
are no longer constant, but depend on the state of the environment $H(t)$ at time $t$. We use ${\rm E}_{\xi} \left[ F(\xi) \bigg| H(t) \right]$ to denote the expectation of $F(\xi)$ when $\xi$ is drawn from the distribution associated with the current state $H(t)$, \emph{i.e.} $f_{\pm}(\xi)$ when $H(t) = H_{\pm}$. In Appendix \ref{cont2app} we derive Eq.~(\ref{eqn:sde_two}) as the continuum limit of the discrete process $y_n$. 

As a consequence of the nonlinearity of Eq.~(\ref{eqn:sde_two}) the state variable $y(t)$ will not drift indefinitely when $g(t)$ is fixed for some time interval $t \in [a,b]$. Rather, trajectories will tend to accumulate about the single fixed point of the noise-free system (the case, $\rho (t) \equiv 0$). Importantly, more volatile environments (larger $\epsilon_{\pm}$) correspond to fixed points that are closer to the midline $y = 0$, allowing for more rapid changes in $\sign \left[ y(t) \right]$. The observer's belief about the environmental state is encoded by the log-likelihood ratio,
and changes at a rate related to the frequency of environmental changes.

The nonlinear term in Eq.~\eqref{eqn:sde_two} does not appear in Eq.~\eqref{eqn:sde_DD}. It serves to discount older evidence by a factor determined by environmental volatility, \emph{i.e.} the frequencies of changes in environmental states, $\epsilon_{\pm}$. In previous work
such discounting was modeled heuristically by a linear term~\cite{usher01,sugrue04,smith04}, however our derivation shows that the resulting Ornstein-Uhlenbeck (OU) process is only an approximation of an optimal observer's evidence accumulation process. 

\subsection{Equal switching rates between two states} When $\epsilon:=\epsilon_+ = \epsilon_-$,  the frequencies of switches between states are equal. Eq.~\eqref{eqn:sde_two} then becomes
\begin{align}\label{eqn:sde_sym}
\d y & = g(t) \d t -2 \epsilon\sinh(y) \d t + \rho(t) \d W_t.
\end{align}
The steepness of the function $\sinh (y)$ at large values of $y$ ensures that evidence is discounted more rapidly for large log-likelihood ratios than for small ones (Fig. \ref{fig:fig4}{\bf F}, below). As a result, evidence builds up faster when $y$ is closer to zero, \emph{i.e.} the observer is more uncertain. If we rescale time using $\tau=\epsilon t$, the rate of switches between environmental states is unity. We obtain an equation for $y_{\tau} : = y(\tau/ \epsilon)$:
\begin{align}\label{eqn:sde_sym_rescaled}
\d y_{\tau} & = \left [ \tilde{g}(\tau) \right] \d\tau -2\sinh(y_{\tau})\d\tau +\left[ \tilde{\rho}(\tau) \right] \d W_{\tau},
\end{align}
where $\tilde{g}(\tau) := g(t)/\epsilon = g(\tau / \epsilon)/\epsilon$ and $\tilde{\rho}(\tau) := \rho(\tau / \epsilon)/\sqrt{\epsilon}.$ 
Recall that $g(t)$ is the rate of evidence accumulation in the present state, and $\epsilon^{-1}$ is the average time
spent in each state.  Hence, $\tilde{g}(\tau) =  g(t) / \epsilon $ can be interpreted as the \emph{information gained over an average duration of the present environmental state}. 

When observations follow Gaussian distributions, $f_\pm\sim \mathcal{N}(\pm \mu,\sigma^2)$, then $g(t)=\pm 2\mu^2 / \sigma^2$, $\rho=2 \mu / \sigma$, and 
\begin{align}\label{eqn:sde_sym_rescaled_normal}
\d y_{\tau} & = \sign[\tilde{g}(\tau)] m \d\tau -2\sinh(y_{\tau})\d\tau +\sqrt{2m} \; \d W_{\tau},
\end{align}
where $m= 2\mu^2 / (\sigma^2\epsilon)$. Thus, the behavior of this system is completely determined by the single parameter $m$, the information gain over an average environmental epoch. 




\begin{figure}
\centering
\includegraphics[width=3.3in]{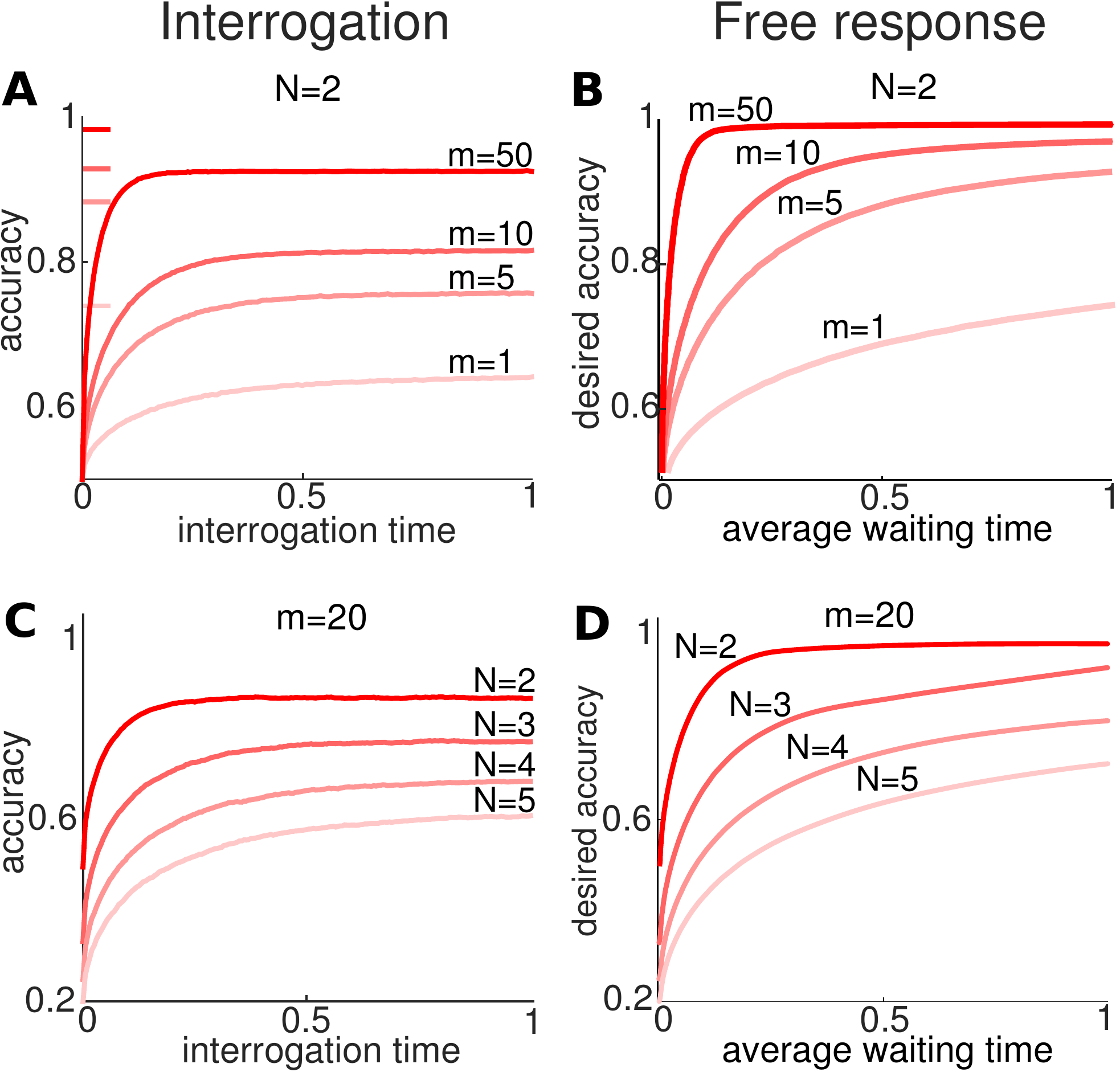}
 \caption{ Dependence of the probability of the correct response (accuracy) on normalized information gain, $m,$ in a symmetric environment.  ({\bf A}) Accuracy in an interrogation protocol increases with $m$ and interrogation time, $t,$ but saturates. Horizontal bars on left indicate the accuracy when the environment is in a single state for a long time, as in Eq.~(\ref{E:stat}). ({\bf B}) When the observer responds freely accuracy is similar, but saturates at 1. The increase in accuracy with waiting time is exceedingly slow for low $m$.  
 We fix $\epsilon_{ij}  \equiv \epsilon$ for all $i \neq j$, $g_i \equiv g$, and set $m = g/\epsilon \equiv 20$.  ({\bf C}) Accuracy in an interrogation protocol decreases with the number of alternatives $N$ (See Section \ref{smultialt} for $N>2$), saturating at ever lower levels. ({\bf D}) The free response protocol results in similar behavior, but the accuracy saturates at 1. The increase in accuracy with waiting time is exceedingly slow for higher numbers of alternatives $N$.
 }
\label{fig:fig3}
\end{figure}

We now analyze the results of two decision-making processes that utilize the log-likelihood ratio. Under the \emph{interrogation protocol}, the observer  waits until a given time $\tau=T$ and reports $\sign \left[ y_{\tau}(T) \right] = \pm 1$. Under the  \emph{free response protocol},  we assume that the observer uses a predetermined threshold, $\theta,$ waits until time $\tau^*$ at which the decision variable meets this threshold, $|y_{\tau}(\tau^*)| =  \theta$, and then reports $\sign \left[ y_{\tau}(\tau^*) \right]$. For Eq.~(\ref{eqn:sde_sym_rescaled_normal}), the probabilities of a correct response (accuracy) under  both interrogation (Fig. \ref{fig:fig3}{\bf A}) and free response (Fig. \ref{fig:fig3}{\bf B}) protocols increase with  $m$. When an optimal observer is interrogated about the
state of the environment at time $T$,  the answer is determined by the sign of the log-likelihood ratio, $y_{\tau}$. Since observers discount old evidence at a rate increasing with $1/m$, decisions are effectively based on a fixed amount of evidence, and accuracy saturates at a value smaller than 1 (Fig. \ref{fig:fig3}{\bf A}). On the other hand, accuracy arbitrarily close to 1 can be obtained in the free response protocol by increasing the threshold $\theta$ (Fig. \ref{fig:fig3}{\bf B}). Equations for the case of multiple alternatives ($N>2$) are provided in Section \ref{smultialt}, and increasing $N$ decreases accuracy for a fixed decision time (Fig. \ref{fig:fig3}{\bf C},{\bf D}).

If the environment in Eq.~(\ref{eqn:sde_sym_rescaled_normal}) remains in a single state for a long time, the log-likelihood ratio, $y_{\tau}$, approaches a stationary distribution,
\begin{equation} \label{E:stat} 
\mathcal{S}_{\pm}(y_{\tau})=K \exp \left(\pm y_{\tau}-\frac{2 \cosh(y_{\tau})}{m}\right), \hspace{3mm} \tilde{H}(\tau) = H_{\pm},
\end{equation}
where $\tilde{H}(\tau) := H(\tau/ \epsilon)$ and $K$ is a normalization constant. Details on finding the stationary density of the Fokker-Planck equation associated with a nonlinear SDE such as Eq.~(\ref{eqn:sde_sym_rescaled_normal}) can be found in Ch.5 of \cite{gardiner04}. The distribution, Eq.~(\ref{E:stat}), is concentrated around $\bar{y}_{\tau \pm} = \pm \sinh^{-1} \frac{m}{2}$,  the fixed points of the deterministic counterpart of Eq.~\eqref{eqn:sde_sym_rescaled_normal} obtained by setting $W_{\tau} \equiv 0$. Since old evidence is continuously discounted, 
the belief of an optimal observer tends to saturate.  In contrast, no stationary distribution exists when $\epsilon =  0$, and the environment is static:
 Aggregating new evidence then always tends to increase an optimal observer's belief in one of the choices.

 Since $\mathcal{S}_{\pm}(y)$ is obtained by assuming that 
the environment is trapped in a single state for an extended time, $\int_{0}^\infty \mathcal{S}_+(y)\d y=\int_{-\infty}^0 \mathcal{S}_-(y)\d y$ provides an upper bound on the accuracy (Fig. \ref{fig:fig3}{\bf A}). To achieve accuracy $a$ in the free response protocol (Fig. \ref{fig:fig3}{\bf B}), we require $|y_{\tau}|\geq \ln \frac{a}{1-a}$ \cite{bogacz06}. While the threshold $\theta =  \ln \frac{a}{1-a}$ that  $y_{\tau}(\tau)$ must cross to obtain a specific accuracy, $a$, does not change with $m$, the time to reach this threshold increases steeply  with $a$ and decreases with $m$. This is partly due to the fact that for smaller $m$, environmental switches are rapid, causing frequent changes in the drift of $y_{\tau}(\tau)$ and keeping it close to the midline $y_{\tau} =0$.

\subsection{Linear approximation of the SDE}

\begin{figure}
\centering
\includegraphics[width=4.5in]{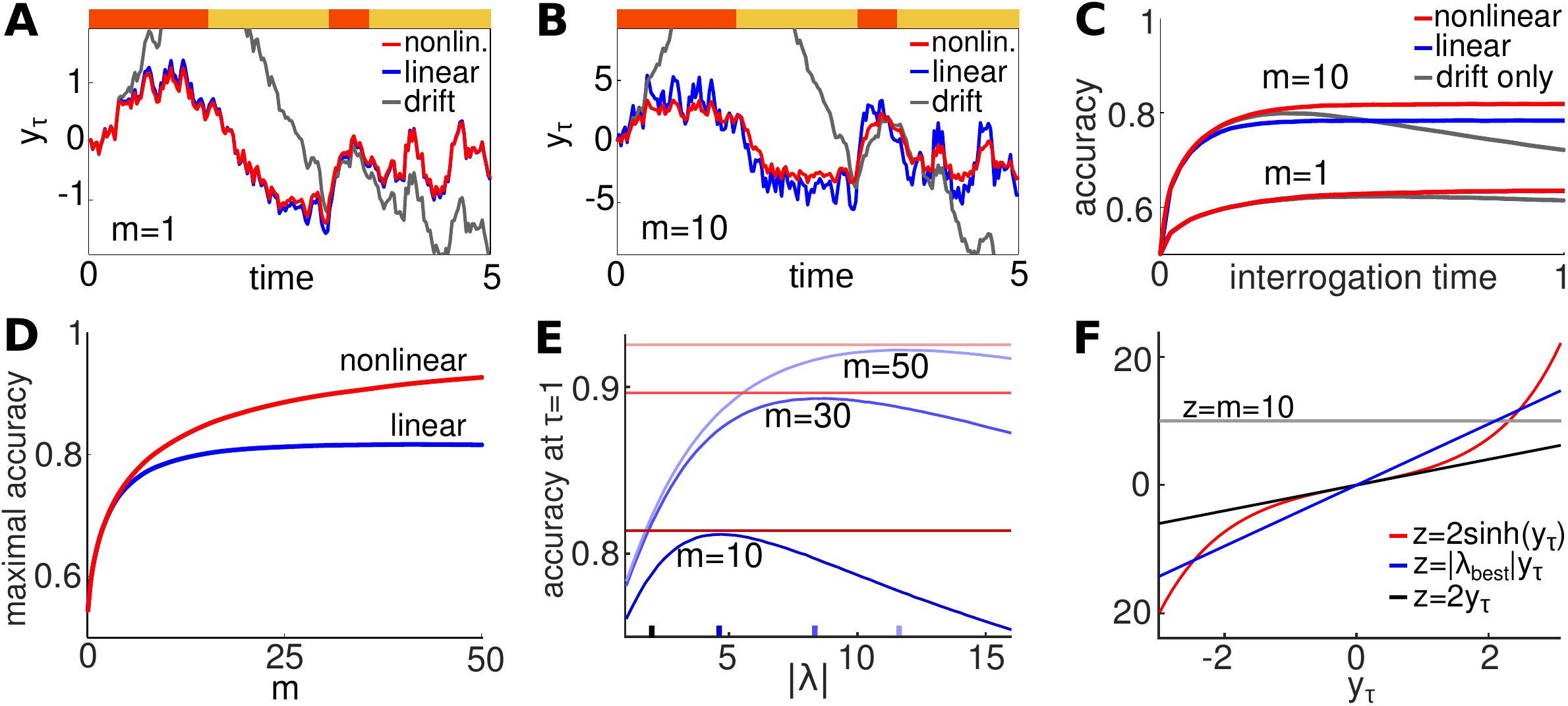}
 \caption{Closest linear approximations of the nonlinear SDE, Eq.~(\ref{eqn:sde_sym_rescaled_normal}). ({\bf A},{\bf B}) Single realizations of the nonlinear Eq.~(\ref{eqn:sde_sym_rescaled_normal}), linear approximation Eq.~(\ref{eq:linapprox}), and corresponding drift-diffusion model $\d y_{\tau} = \sign[\tilde{g}(\tau)] m \d \tau + \sqrt{2m} \d W_{\tau},$ in ({\bf A}) a quickly changing environment ($m=1$), and ({\bf B}) a slowly changing environment ($m=10$). We used the same realizations of drift $\tilde{g}(\tau),$ and noise $W_{\tau}$ for all models. ({\bf C}) In the interrogation protocol, accuracy increases faster in the nonlinear Eq.~(\ref{eqn:sde_sym_rescaled_normal}) than in the linear approximation Eq.~(\ref{eq:linapprox}). Accuracy eventually decreases in the drift model since all evidence is weighted equally across time. ({\bf D}) In the limit $t\to \infty$, accuracy saturates below unity in both the nonlinear model and linear approximation. The linear model discounts evidence sub-optimally,  and hence performs worse.  ({\bf E}) Accuracy under the interrogation protocol with stopping time $\tau=1$ for the linear model, Eq.~(\ref{eq:linapprox}), with leak $\lambda$ (blue ticks: $\lambda_{\text{best}}$, black tick: $\lambda=-2$). Optimal $\lambda$-values for the linear approximation (blue curves) result in accuracy that is very close to that of the optimal nonlinear model, Eq.~(\ref{eqn:sde_sym_rescaled_normal}) (red lines). ({\bf F}) Plot of $2\sinh (y_{\tau})$ demonstrating two possible linear approximations: most accurate linear approximation from panel {\bf E}   (blue), linearization of $2 \sinh( y_{\tau})$ at the origin (black).
 }
\label{fig:fig4}
\end{figure}

An advantage of  Eq.~(\ref{eqn:sde_two}) is that it is amenable to standard methods of stochastic analysis. We can find  an accurate piecewise linear approximation to Eq.~(\ref{eqn:sde_two}), although, for simplicity, we focus on Eq.~(\ref{eqn:sde_sym_rescaled_normal}). 
The  piecewise OU process that models an observer that linearly discounts 
evidence has the form  
\begin{align} 
\d y_{\tau} = b ( \sign [\tilde{g}(\tau)] m  d \tau + \sqrt{2m} \d W_{\tau} ) + \lambda y_{\tau} {\rm d} \tau. \label{eq:linapprox}
\end{align}
For Eq.~(\ref{eq:linapprox}) to be the continuum limit of a linear log-likelihood update process, the drift and diffusion need to be co-scaled by the common parameter $b$. We begin by focusing on a linear approximation of Eq.~(\ref{eqn:sde_sym_rescaled_normal}) with the same equilibria and local stability, obtained by setting
$\lambda = - \sqrt{m^2 + 4}$ and
$b = \sqrt{1+\frac{4}{m^2}} \sinh^{-1} \frac{m}{2}$. Individual realizations of Eq.~(\ref{eq:linapprox}) and  Eq.~(\ref{eqn:sde_sym_rescaled_normal}) agree in quickly changing environments (Fig. \ref{fig:fig4}{\bf A}, $m=1$), but are less similar in slowly changing environments (Fig. \ref{fig:fig4}{\bf B}, $m=10$; see also panel {\bf C}). Thus, as observer performance improves, the nonlinear term in Eq.~\eqref{eqn:sde_sym_rescaled_normal} becomes more important. Note that the corresponding drift-diffusion model, $\d y_{\tau} = \sign \left[ \tilde{g}(\tau) \right] m \d \tau + \sqrt{2 m} \d W_{\tau}$, is qualitatively different as it lacks a restorative leak term. This difference becomes more pronounced as $m$ increases (Fig. \ref{fig:fig4}{\bf C}).

Eq.~(\ref{eq:linapprox}) can be integrated explicitly using standard methods in stochastic calculus~\cite{gardiner04}. Furthermore, the accuracies of both systems saturate to a value smaller than 1 in the interrogation protocol as the interrogation time increases (Fig. \ref{fig:fig4}{\bf C}). 

This linearized approximation can differ considerably from the full nonlinear model. 
For instance, in the interrogation protocol the performance of an ideal observer modeled by  Eq.~(\ref{eqn:sde_sym_rescaled_normal})  increases with interrogation time (Fig. \ref{fig:fig4}{\bf D}),  and accuracy approaches 1 as $m$ diverges. In contrast, the accuracy of an observer that discounts evidence linearly limits to a value below 1 as $m$ diverges. Indeed, this can be seen by employing the quasi-steady state approximation (fixing $\sign [ \tilde{g}(\tau) ] = 1$), and computing $\int_{0}^{\infty}\mathcal{T}(y)dy $, where $\mathcal{T}(y)$ is the steady state distribution of the OU process  given in Eq.~\eqref{eq:linapprox} with $\lambda = - \sqrt{m^2 + 4}$ and $b = \sqrt{1+\frac{4}{m^2}} \sinh^{-1} \frac{m}{2}$, 
to obtain $$U_m: = \int_{0}^{\infty}\mathcal{T}(y)dy =\frac{1}{2}+\frac{1}{2}{\rm erf}\left(\sqrt{\frac{m}{2 \sqrt{m^2+4}}}\right),$$ and $\lim_{m\rightarrow \infty}U_m = \frac{1}{2} + \frac{1}{2} {\rm erf} \frac{1}{\sqrt{2}} \approx 0.84 <1$. 

 In general, there is a family of linear approximations to Eq.~(\ref{eqn:sde_sym_rescaled_normal}) given by Eq.~\eqref{eq:linapprox}, where $\lambda \in (-\infty,0]$. However, the choice of $\lambda$ depends on the way we measure the quality of the approximation. For example, we may want to maximize decision accuracy under the interrogation policy with a specific stopping time, or maximize accuracy under the free response policy. In general, we need numerical optimization methods to identify the $\lambda$ that provides the best linear approximation. Without loss of generality, we can fix $b \equiv 1$ in Eq.~\eqref{eq:linapprox}, since the rescaling $z_{\tau} = y_{\tau}/b$ preserves $\sign \left[ z_{\tau} \right] = \sign \left[ y_{\tau} \right]$ and eliminates $b$. Thus, we need only study the system $\d y_{\tau} =\sign \left[ \widetilde{g}(\tau) \right] m \d \tau + \sqrt{2 m} \, \d W_{\tau} + \lambda y_{\tau} \d \tau$.  For a given $m$, there is a single value, $\lambda = \lambda_{\text{best}},$ that maximizes the accuracy of decisions after an interrogation at $\tau=1$ (Fig. \ref{fig:fig4}{\bf E}). However, there is a different $\lambda_{\text{best}}$ for each value of $m$. Interestingly, the best linear approximation has accuracy  close to that of the nonlinear system. We note that the linear approximation at the origin ($\lambda=-2$, see also Fig. \ref{fig:fig4}{\bf F}) did not perform well. Since  the accuracy has saturated at $\tau=1$ (Fig. \ref{fig:fig4}{\bf C}), the optimal value of $\lambda$ will not change significantly for larger interrogation times.
 
Similarly, $\lambda_{\text{best}}$ will change with different thresholds under the free response protocol, or with other  measures of performance, such as reward rate~\cite{bogacz06}. In contrast, the nonlinear model given by Eq.~(\ref{eqn:sde_sym_rescaled_normal}), reflects the log-likelihood ratio exactly.  Therefore, we can use this
single model for any decision that can be made optimally using the log-likelihood ratio.


\section{Multiple alternatives in a changing environment} \label{smultialt}
\begin{figure}
\centering
\includegraphics[width=3.3in]{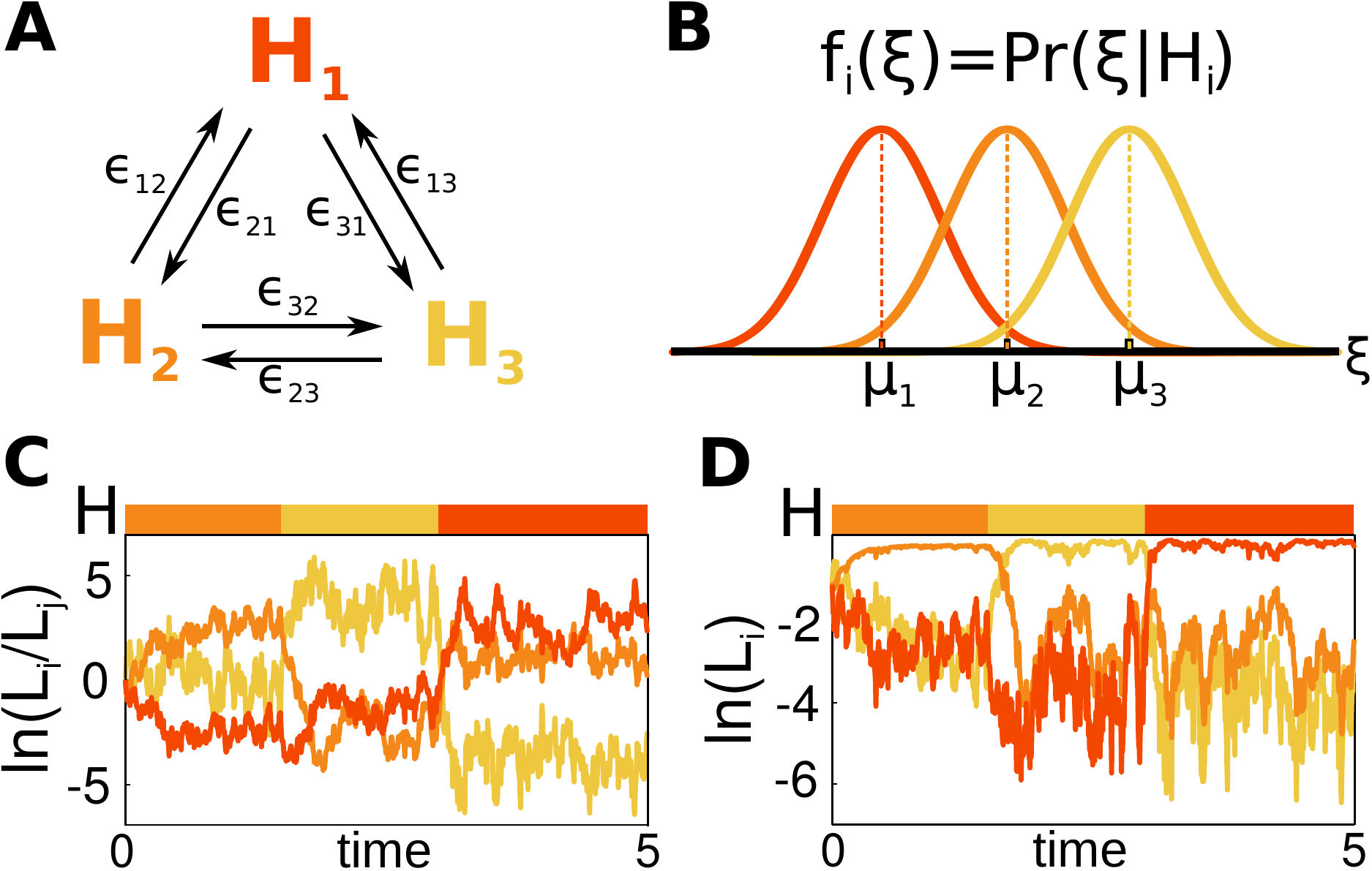}
\vspace{-.2cm}
\caption{Evidence accumulation with multiple choices in a changing environment. ({\bf A}) The environment switches between $N$ states (here $N=3$). ({\bf B}) Distributions $f_{i}(\xi) = {\rm Pr}(\xi | H_i)$ describing the probability of  observation $\xi$ in  environmental state $H_i$ (here $N=3$). 
({\bf C,D}) Realization of the log-likelihood ratios (panel \textbf{C}): $\ln(L_1/L_2)$, $\ln(L_2/L_3)$, $\ln(L_3/L_1)$, and 
(panel \textbf{D}): $\ln L_1$, $\ln L_2$, $\ln L_3$. }
\label{fig:fig5}
\end{figure}

We next extend our analysis of evidence accumulation in changing environments to the case of multiple alternatives. With multiple environmental states, $H_i$ ($i=1,...,N$), the optimal observer computes the present probability of each state (Fig~\ref{fig:fig5}{\bf A}) from a sequence of measurements, $\xi_{1:n}.$  Measurements have probability
$f_{i}(\xi_n):={\rm Pr}(\xi_n | H_i)$ dependent on the states $H_i$~\cite{bogacz07b,beck08} (Fig~\ref{fig:fig5}{\bf B}).
We assume that the state of the environment, $H(t),$ changes as a memoryless process.  A change from state $j$ to $i$ between two measurements occurs with probability $\epsilon_{ij} \Delta t = {\rm Pr}(H(t_n) = H_i | H(t_{n-1}) = H_j )$ for $i\neq j$, and 
${\rm Pr}(H(t_n) = H_i | H(t_{n-1}) = H_i )=1-\sum_{j\neq i}\Delta t\epsilon_{ji}  $ 
(Fig~\ref{fig:fig5}{\bf A}).

We again use sequential analysis to obtain the probabilities $L_{n,i}  = {\rm Pr}(H(t) = H_i | \xi_{1:n}) $ that the environment is in state $H_i$ given observations $\xi_{1:n}$. The index that maximizes the posterior probability, $\widehat{\imath} = {\rm argmax}_i \; L_{n,i},$ corresponds to the most probable state, given the observations $ \xi_{1:n}$. Following the approach above, we obtain (See Appendix \ref{llmultiapp}):
\begin{equation*}\label{eqn:general_disc_many}
L_{n,i}  = \frac{{\rm Pr}(\xi_{1:n-1})}{{\rm Pr}(\xi_{1:n})} f_i(\xi_n)
\left(
\left(1-\sum _{j\neq i} \Delta t \epsilon_{ji}\right)L_{n-1,i} +
\sum _{j\neq i} \Delta t \epsilon_{ij}L_{n-1,j} 
\right). 
\end{equation*}
Again after taking logarithms, $x_{n,i} = \ln L_{n,i}$,  we can approximate the discrete stochastic process in Eq.~(\ref{eqn:general_disc_many}), with an SDE:
\begin{align}\label{eqn:vecsde}
\d \x & = \g(t) \d t +\Lambda (t) \d \W_t + K(\x) \d t,
\end{align}
where the drift has components $g_i(t) = \frac{1}{\Delta t}{\rm E}_{\xi} \left[\ln f_i(\xi) | H(t) \right]$, $\Lambda (t) \Lambda(t)^T = \Sigma (t)$  with entries $\Sigma_{ij}=\frac{1}{\Delta t} \text{Cov}_{\xi}[\ln f_i(\xi),\ln f_j(\xi)|H(t)]$, components of $\W_t$ are independent Wiener processes, and $K_i(\x) = \sum_{j \neq i} (\epsilon_{ij} \e^{x_j - x_i} - \epsilon_{ji})$. 
The drift $g_i$ is maximized in environmental state $H_i$ (Fig~\ref{fig:fig5}{\bf C},{\bf D}). 

We can recover the  case of two alternatives by setting $N=2$ and  exchanging the numbers in Eq.~(\ref{eqn:vecsde}) with $\pm$ to obtain the approximating SDEs:
\begin{align} \label{eqn:sde_two_pm}
\d x_{\pm} & = [g_{\pm}(t) + \left( \epsilon_{\mp}e^{x_{\mp} - x_{\pm}}-\epsilon_{\pm} \right)] \d t + \d W_{\pm},  
\end{align}
where $\langle W_i W_j \rangle = \Sigma_{ij}(t) \cdot t$ for $i,j \in \{ +, - \}$.  We obtain Eq.~(\ref{eqn:sde_two}) by setting  $y:=x_+-x_-$. Note that since $x_{\pm} = \ln (L_{\pm})$ are the log-likelihoods, $y:=x_+ - x_- = \ln (L_+/ L_-)$ is the log-likelihood ratio. Analogous expressions for the log-likelihood ratios $y_{ij} = \ln (L_i/L_j)$ are derived in Appendix \ref{llrmultiapp}. The matrix of these log-likelihood ratios quantifies how much more likely one alternative is compared to others (\emph{e.g.}, Fig. \ref{fig:fig5}{\bf C}) \cite{draglia99}.

\begin{figure}
\centering
  \includegraphics[width=3.3in]{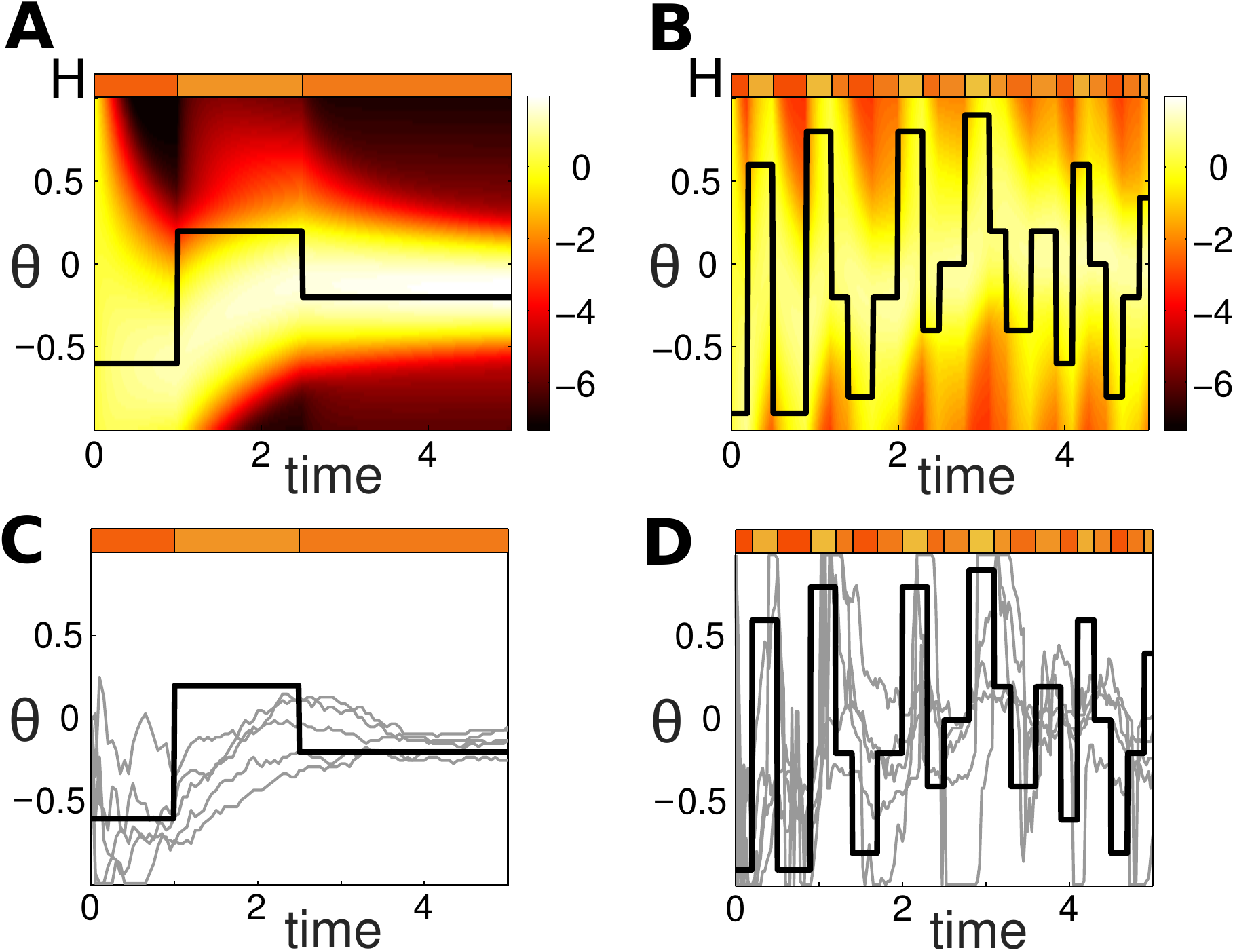}
  \vspace{-.2cm}
 \caption{Evidence accumulation with a continuum of choices. The observer  infers the state of the environment, $H_\theta$, where $\theta \in [-1,1],$ and the state changes at discrete points in time.   ({\bf A})   In slowly changing environments, the distribution of the log probabilities, $x_\theta,$
 can nearly equilibrate between switches (Solid line represents the true state of the environment at time $t$. For clarity, we show results of simulations without noise, $\widehat{W}_{\theta} \equiv 0$). ({\bf B}) In quickly changing environments, the distribution does not have time to equilibrate between switches.  ({\bf C}) In slowly changing environments, the most probable state of the environment, $\widehat{\theta}(t)  = {\rm argmax}_{\theta} x_\theta(t)$ (thin lines), fluctuates around the true value (thick line).  ({\bf D}) In quickly changing environments, $\widehat{\theta}(t)$ fluctuates more widely, as it is in a transient state much of the time.
 \vspace{-3mm} }
\label{fig:fig6}
\end{figure}


\section{A continuum of states in a changing environment} Lastly, we consider the case of  a continuum of possible environmental states. 
This provides a tractable model for recent experiments with observers who infer the location of a hidden, intermittently moving 
target from noisy observations. Evidence suggests that humans update their beliefs quickly and near optimally
when observations indicate that the target has moved~\cite{mcguire14}.

Suppose the environmental state, $H(t),$  intermittently switches between a continuum of possible states, $H_{\theta},$ where $\theta \in [a,b]$. An observer again computes the probabilities 
 of each state from observations, $\xi_{1:n}$, with distributions $f_{\theta} (\xi_n) : = {\rm Pr}(\xi_n|H_{\theta})$. The environment switches from state $\theta'$ to state $\theta$ between observations with transition probabilities $\epsilon_{\theta \theta'}  \d \theta \Delta t :=  {\rm Pr} (H(t_n) = H_{\theta} | H(t_{n-1}) = H_{\theta'})$  for $\theta\neq\theta'$, and ${\rm Pr} (H(t_n) = H_{\theta} | H(t_{n-1}) = H_{\theta})=1-\int_{a}^b \Delta t \epsilon_{\theta'\theta}d \theta'$ (See Appendix \ref{contapp} for details). 
From Eq.~\eqref{eqn:general_disc_many} the expression for the probabilities $L_{n,\theta} = {\rm Pr}(H(t_n) =  H_{\theta}|\xi_{1:n})$ is derived in Appendix \ref{contapp}, yielding: 
\begin{equation*} 
L_{n,\theta}  = \frac{{\rm Pr}(\xi_{1:n-1})}{{\rm Pr}(\xi_{1:n})} f_\theta(\xi_n) 
 \left( \left( 1-\int_a^b\Delta t \epsilon_{\theta'\theta}d\theta' \right) L_{n-1,\theta}+ 
\int_a^b \Delta t \epsilon_{\theta\theta'}L_{n-1,\theta'}d\theta' 
\right).
\end{equation*}
We again approximate the logarithms of the probabilities, $\ln L_{n,\theta}$, by a temporally continuous process,
\begin{align}\label{eqn:sde_cont}
\d x_{\theta}(t) & = g_{\theta}(t) \d t + \d \widehat{W}_{\theta}(t) +  K_\theta(x(t)) \d t ,
\end{align}
where, $x=\left(x_\theta \right)_{\theta\in[a,b]}$, $g_\theta(t) = \frac{1}{\Delta t} {\rm E}_\xi \left[ \ln f_\theta (\xi) | H(t) \right]$, $\widehat{W}_{\theta}$ is a spatiotemporal noise term with mean zero and covariance function given by
\begin{align*}
\Sigma_{\theta\theta'}(t)
= \frac{1}{\Delta t} \text{Cov}_\xi\left[ \ln f_{\theta}(\xi), \ln f_{\theta'}(\xi)| H(t) \right],
\end{align*}
and $ K_\theta(x)=  \int_a^b( \epsilon_{\theta\theta'} e^{x_{\theta'}-x_{\theta\phantom{'}}} -\epsilon_{\theta'\theta})d\theta'$  is an interaction term describing the discounting process.

The drift $g_\theta(t)$ is maximal when $\theta$ agrees with the present environmental state. 
The most likely state, given observations up to time $t$, is  $\widehat{\theta}(t)  = {\rm argmax}_{\theta} x_\theta(t)$. 

In slowly changing environments, the log probability $x_{\theta} (t)$ \emph{nearly} equilibrates to a distribution with a well-defined peak between environmental switches (Fig. \ref{fig:fig6}{\bf A}). This does not occur in quickly changing environments (Fig. \ref{fig:fig6}{\bf B}). However, each logarithm, $x_\theta(t)$  approaches a stationary distribution 
if the  environmental state remains fixed for a long time. The term $K_{\theta}(x)$ in Eq.~(\ref{eqn:sde_cont}) causes rapid departure from this quasi-stationary density when the environment changes, a mechanism proposed in \cite{mcguire14}. 

Even when the environment is stationary for a long time, noise in the observations stochastically perturbs the log probabilities, $x_{\theta} (t)$, over the environmental states. This leads to fluctuations in the estimate $\widehat{\theta}(t)$ of the most probable alternative (Fig. \ref{fig:fig6}{\bf C},{\bf D}). Thus, as opposed to the case of a discrete space of $N$ alternatives, the observer's estimate of the most probable choice will change continuously, fluctuating about the continuum of possible alternatives. Unless changes are too rapid, the peak of the log probability distribution,  $\widehat{\theta}(t)$, fluctuates around the true environmental state, and tracks abrupt changes in $H_{\theta}(t)$.  This is in line with recent observations in human behavioral data~\cite{mcguire14,glaze15}.


\section{A neural implementation of an optimal observer} 
Previous neural models of decision making typically relied on mutually inhibitory neural networks \cite{usher01,wang02,mcmillen06,bogacz07}, with each population representing one alternative. 
These models match the recorded neural activity and responses of monkeys performing two-alternative forced-choice decision tasks, where single trial stimuli have stationary statistics~\cite{gold07}. 
Even when reward rates are varied across trials, animals can adjust their behavior near-optimally from trial-to-trial in ways that are well captured by mutually inhibitory models~\cite{Feng09}. Interestingly, these networks also provide a plausible model of decision-making in house-hunting honeybee swarms~\cite{Pais13}.  In previous studies, it has been shown that a single fixed point can be stabilized in linear population models, as long as the strength of mutual inhibition is weaker than the leak of individual populations~\cite{usher01,bogacz06,bogacz07}. As we will show, a complementary approach in linear population models is to consider a mutually excitatory network, with arbitrary leak in individual populations. As with the linear
approximations discussed above,  such models perform suboptimal inference in changing environments, 
but can approach the performance of the ideal nonlinear discounting process given by Eq.~(\ref{eqn:sde_two}).

\begin{figure}
\centering
\includegraphics[width=3in]{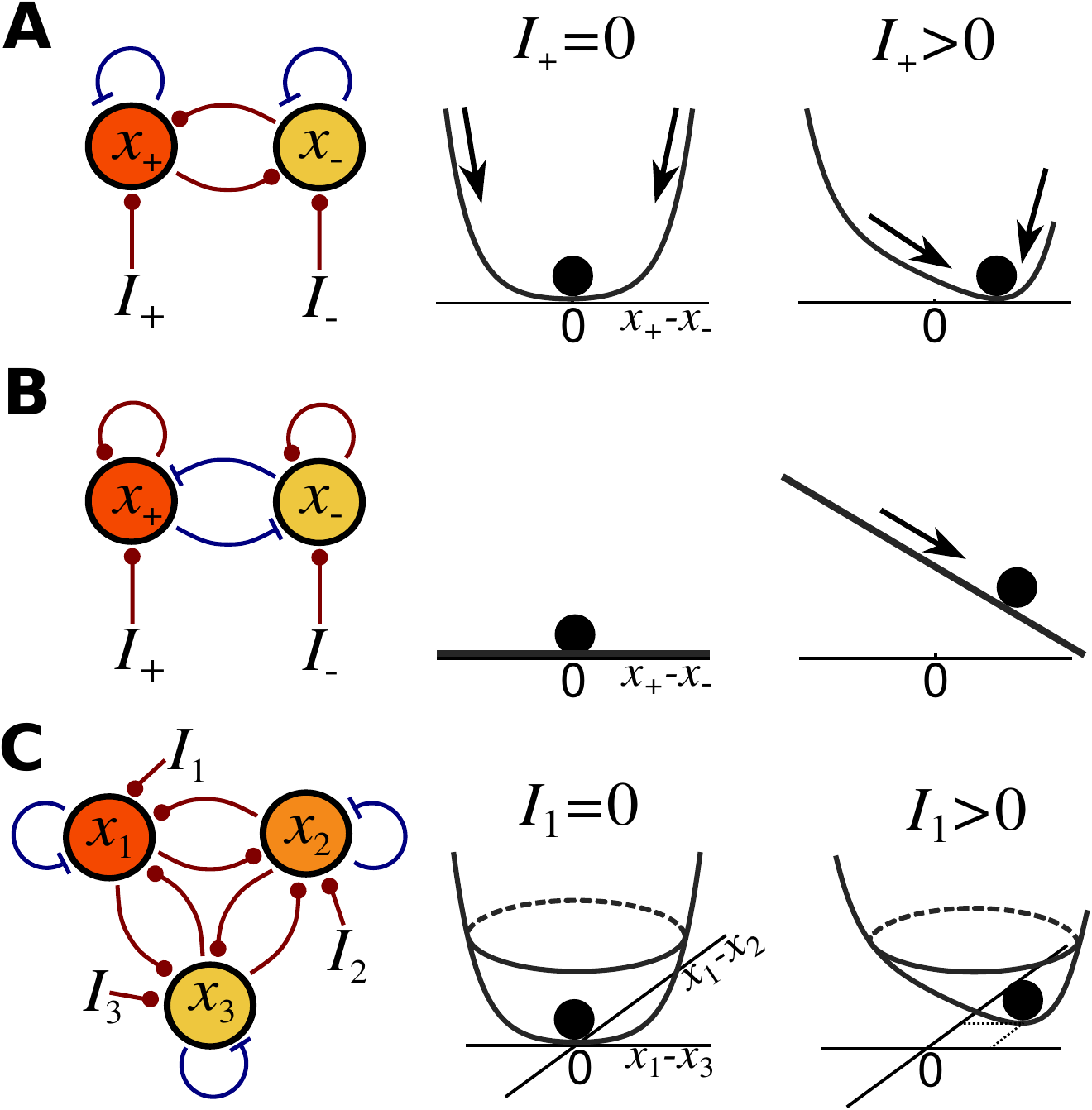}
 \caption{Neural population models of evidence accumulation. ({\bf A}) Two populations $u_{\pm}$ receive a fluctuating stimulus with mean $I_{\pm}$; they are mutually coupled by excitation (circles) and locally affected by inhibition/leak (flat ends). When $I_+>0$, the fixed point of the system has coordinates satisfying $x_+>x_-$ as shown in the plots of the associated potentials. ({\bf B}) Taking $\epsilon_{\pm} \to 0$ in Eq.~(\ref{neurpop}) generates a mutually inhibitory network that perfectly integrates inputs $I_{\pm}$ and has a flat potential function. ({\bf C}) With $N=3$ alternatives, three populations coupled by mutual excitation can still optimally integrate the inputs $I_{1,2,3}$, rapidly switching between the fixed point of the system in response to environmental changes. \vspace{-.3cm}}
\label{fig:fig7}
\end{figure}

Optimal inference in dynamic environments with two states, $H_+$ and $H_-,$ can be performed by mutually excitatory nonlinear neural populations with activities (firing rates) $r_+$ and $r_-$ evolving according to:
\begin{subequations} \label{neurpop}
\begin{align}
\d r_+ &= \left[ I_+(t) - \alpha r_+ + F_+(r_--r_+) \right] \d t +  \d W_+,  \\
\d r_- &= \left[ I_-(t) - \alpha r_- + F_-(r_+-r_-) \right] \d t +  \d W_- , 
\end{align}
\end{subequations}
where the transfer functions are $F_{\pm}(x) = -\alpha x/2 + \epsilon_{\mp}\e^{x} - \epsilon_{\pm}$,  the mean input $I_{\pm} (t) = I_{\pm}^0$ when $H(t) = H_{\pm}$ and vanishes otherwise,
$W_{\pm}$ are Wiener processes representing the variability in the input signal with covariance defined as in Eq.~(\ref{eqn:sde_two_pm}) (See Appendix \ref{llmultiapp}). Thus, $I_{\pm}(t) \d t + \d W_{\pm}$ represents the total input to population $r_{\pm}$. When $\alpha>0$ and sufficiently small,  population activities are modulated by self-inhibition/leak, and mutual excitation (Fig. \ref{fig:fig7}{\bf A}). The parameter $\alpha$ determines the leak in each individual population, which depends on both the time constants and recurrent architecture of the local network \cite{wang02}. The difference $y = r_+ - r_-$ evolves according to the SDE for the log-likelihood ratio, Eq.~(\ref{eqn:sde_two}). In the limit of a stationary environment, $\epsilon_{\pm} \to 0$, we obtain a linear integrator $\d r_{\pm} = \left[ I_{\pm} \d t + \d W_{\pm} \right] - \alpha (r_+ + r_-) \d t/2$, as in previous studies~\cite{bogacz06,mcmillen06}. 

To show that the populations mutually excite each other, we set $W_+ = W_- = 0,$ and study the dynamics in the vicinity of the fixed points of Eqs.~\eqref{neurpop}. When the environment has not changed for a long time, Eq.~(\ref{neurpop}) approaches a fixed point $(\bar{r}_+,\bar{r}_-)$ with 
\begin{align*}
(\bar{r}_+,\bar{r}_-) =
\left(\frac{I_+^0 + \epsilon_- {\rm e}^{-\bar{y}_+}-\epsilon_+}{\alpha}+\frac{\bar{y}_+}{2}, 
\frac{\epsilon_+{\rm e}^{\bar{y}_+}-\epsilon_-}{\alpha}-\frac{\bar{y}_+}{2} \right)
,
\end{align*}  
when $I_+(t) = I_+^0$ and $I_-(t) = 0$ and
\begin{align*}
(\bar{r}_+,\bar{r}_-) =
\left(\frac{\epsilon_- {\rm e}^{-\bar{y}_-}-\epsilon_+}{\alpha}+\frac{\bar{y}_-}{2}, 
\frac{I_-^0 + \epsilon_+{\rm e}^{\bar{y}_-}-\epsilon_-}{\alpha}-\frac{\bar{y}_-}{2} \right)
,
\end{align*}
when $I_+(t) = 0$ and $I_-(t) \equiv I_-^0$, where
\begin{align*}
 \bar{y}_{\pm} & = \ln \left[ \frac{\pm I_{\pm}^0 + \epsilon_- - \epsilon_+}{2\epsilon_+} + \sqrt{\frac{(\pm I_{\pm}^0 + \epsilon_- - \epsilon_+)^2}{4\epsilon_+^2}+\frac{\epsilon_-}{\epsilon_+}} \right].
\end{align*} 
Note that by increasing (decreasing) $\alpha$, the fixed points $(\bar{r}_+, \bar{r}_-)$ move closer to (farther from) the origin $(0,0)$. To determine the sign of the coupling near these fixed points, note that the Jacobian matrix of $(F_+,F_-)$ has the form: 
\begin{align}
J(r_+,r_-)=\left[\begin{array}{rr}
\alpha/2-\epsilon_- {\rm e}^{r_- - r_+} & -\alpha/2+\epsilon_- {\rm e}^{r_- - r_+}\\
-\alpha/2+\epsilon_+ {\rm e}^{r_+ - r_-} & \alpha/2-\epsilon_+ {\rm e}^{r_+ - r_-}
\end{array}\right].  \label{neurjacob}
\end{align}
For $\epsilon_\pm>0$, taking $\alpha<2\min\{ \epsilon_- {\rm e}^{-\bar{y}} ,  \epsilon_+ {\rm e}^{\bar{y}}  \}$ will guarantee that the sign of the Jacobian matrix is $\left[\begin{array}{rr}
- & +\\
+ & -
\end{array}\right]$ in a region that contains the fixed point. This corresponds to a neural network with self-inhibition/leak and mutual excitation illustrated in Fig. \ref{fig:fig7}{\bf A}.

We can compare our results to previous studies of linear connectionist models~\cite{usher01,bogacz06,bogacz07} by deriving a linear rate model that best accumulates evidence in changing environments. To do so, we focus on the best linear approximation of the log-likelihood ratio, given by Eq.~(\ref{eq:linapprox}). We have shown that when the coefficients of the linear models are tuned appropriately, their accuracy is remarkably close to that of the full nonlinear model (Fig. \ref{fig:fig4}{\bf E}). Assuming symmetric switching ($\epsilon_{\pm} \equiv \epsilon$), the following system describes a linear rate model that can be mapped to the linear Eq.~(\ref{eq:linapprox}):
\begin{subequations} \label{neurpoplin}
\begin{align}
\d r_+ &= \left[ I_+(t) - \kappa r_+ + \gamma r_- \right] \d t + \d W_+, \\
\d r_- &= \left[ I_-(t) - \kappa r_- + \gamma r_+ \right] \d t + \d W_-.
\end{align}
\end{subequations}
Here $\kappa >0$ denotes the leak in each population's activity, and $\gamma >0 $ is the strength of mutual excitation between populations. Selecting $I_{\pm}(t) = I_0$ when $H(t) = H_{\pm}$ and zero otherwise, it can be shown that the system will tend to the quasi-equilibria $\D (r_+,r_-) = \frac{I_0}{\kappa^2 - \gamma^2} \cdot (\kappa, \gamma)$ and $\D \frac{I_0}{\kappa^2 - \gamma^2} \cdot (\gamma,\kappa)$ in either case. Stability of these fixed points is given by the nonzero eigenvalue $\lambda = - (\kappa + \gamma)<0$, so these quasi-equilibria are always attractive. Note also that the reduced SDE for the difference $y = r_+ - r_-$ will take the form $\d y = \left[ I_d(t) - (\kappa + \gamma) y \right] \d t + \d W_d$, where $I_d(t) = I_+(t) - I_-(t)$ and $W_d = W_+ - W_-$, which matches the form of Eq.~(\ref{eq:linapprox}).
Thus, in addition to the large leak  in mutually inhibitory networks~\cite{usher01,bogacz06,bogacz07}, linear population networks with mutual excitation possess a stable fixed point for arbitrary leak $\kappa$ and mutual excitation $\gamma$. Any particular decision task has an optimal $\lambda$ in Eq.~(\ref{eq:linapprox}). Thus, a linear neural network could be trained to learn this best evidence discounting rate if supplemented with a plasticity rule that properly tunes the excitation strength $\gamma$.

Returning to the nonlinear model given by Eq.~(\ref{neurpop}), the dynamics is matched to the timescale of the environment determined by $\epsilon_{\pm},$ and solutions approach stationary distributions if input is constant.  The network's dynamics is very sensitive to changes in inputs, a feature absent in population models with winner-take-all dynamics~\cite{wong07}. Even when $\epsilon$ is small, Eq.~(\ref{neurpop}) has a single attracting state determined by the mean inputs $I_{\pm}^0$. We illustrate the response of the model to inputs using potentials (Fig. \ref{fig:fig7}{\bf A}). In contrast to the single attractor of  Eq.~(\ref{neurpop}), mutually inhibitory models can possess a neutrally stable line attractor that integrates inputs ($\epsilon_{\pm} \equiv 0$, Fig. \ref{fig:fig7}{\bf B}) \cite{machens05}.

We can extend our results  to $N>2$. In \cite{beck08}, the reliability of motion information was assumed to vary during a trial, and the optimal model encoded the posterior probability distribution over the possible stimulus space. Here, we assume the true hypothesis, $H(t),$ changes in time. For an arbitrary but finite number of possible alternatives, $\{ H_1, ..., H_N \}$, decisions  can be performed optimally by neural populations $(r_1, ..., r_N)$ coupled by mutual excitation
\begin{align}
\d r_i = \left[ I_i(t) - \alpha r_i + \sum_{j \neq i} F_{ij} (r_j - r_i ) \right] \d t + \d \widehat{W}_i(t),  \label{npmulti}
\end{align}
where the mean input is $I_i (t) = I_i^0$ when $H(t) = H_i$ and $0$ otherwise and the noise vector
$(\d \widehat{W}_1(t), ..., \d \widehat{W}_N(t))^T = \Lambda (t) \d \mathbf{W}_t$ describes input noise with $\Lambda (t)$ defined as in Eq.~(\ref{eqn:vecsde}). Population firing rates are again determined by inhibition/leak within each population and excitation between populations as described by the arguments of the firing rate function
\begin{align*}
F_{ij}(r) = - \alpha r/N + \epsilon_{ij} \e^r - \epsilon_{ji}.
\end{align*}
In this case coupling between populations is again  excitatory (Fig. \ref{fig:fig7}{\bf C}).  

Note that, as in the case of $N=2$ alternatives, taking the limit of Eq.~(\ref{npmulti}) as $\epsilon_{ij} \to 0$, we obtain linear integrators~\cite{mcmillen06}
\begin{align*}
\d r_i = \left[ I_i(t) - \alpha \sum_{j=1}^N r_j/N \right] \d t + \d \widehat{W}_i (t).
\end{align*}

The nonlinear population rate models described by Eq.~(\ref{neurpop}) and Eq.~(\ref{npmulti}) react rapidly, but not instantaneously,  to changes in their inputs. Recent evidence suggests that in monkeys the activity of single neurons in area LIP exhibits jumps, rather than a gradual increase as previously suggested~\cite{Latimer15}. Furthermore, the performance of rats and humans discriminating the direction of auditory click sequences can be optimally fit by  a pulse--accumulating mechanism with zero noise~\cite{brunton13}. 
However, the activity of a population of cells encoding behavior may still ramp upwards or downwards.

\section{Discussion}

We have derived a nonlinear stochastic model of optimal evidence accumulation in changing environments. Importantly, the resulting SDE is not an OU process, as suggested by previous heuristic models \cite{ratcliff78,usher01,smith04}. Rather, an exponential nonlinearity allows for optimal discounting of old evidence, and rapid adjustment of decision variables following environmental changes. As a result, the certainty of an optimal observer tends to saturate, even if the environment happens to 
be stuck in a single state for long periods of time.

We have made several assumptions about the model to simplify the derivations. Our ideal observer is assumed to be aware both of the uncertainty of their own measurements and about the frequency with which the environment changes. A more realistic model would require that a naive observer learn the underlying volatility of the environment. Modeling the case of initially unknown transition rates  leads to hierarchical models that identify the location of change-points~\cite{wilson10}. However, this approach quickly grows in computational complexity, since the probability of change points is determined by accounting for all possible transition histories~\cite{adams07}.
We also assumed that changes in the environment follow a memoryless process. In more general cases, we would
not be able to obtain a recursive equation for the probability 
 of a state. An ideal observer would have to use all previous
observations at each step, rather than integrating the present observation with the posterior probability obtained with the previous observation. This process cannot be approximated by an SDE.

Sequential sampling in dynamic environments with two states has been studied previously in special cases, such as adapting spiking models, capable of responding to environmental changes \cite{deneve08}. Likelihood update procedures have also been proposed for multiple alternative tasks in the limit $\epsilon_{ij} \to 0$~\cite{armitage50,draglia99}, but their dynamics was not analyzed. A related case of a temporally changing context has also been examined~\cite{shvartsman15}.  One important conclusion of our work is that $m=g/\epsilon$, the information gain over the characteristic environmental timescale, is the key parameter determining the model's dynamics and accuracy. It is easy to show that equivalent parameters govern the dynamics of likelihoods of multiple choices. This allows for a straightforward approximation of the nonlinear model by a linear SDE, which can be analyzed fully. 

Models of evidence accumulation
are of interest in disciplines ranging from neuroscience 
and robotics to psychology and economics. They can help us understand how  
decisions are made in cells, animals, ecological groups, and social networks.
We presented a principled derivation of a series of nonlinear stochastic models  amenable to 
stochastic analysis, and have used quasi-static approximations, 
first passage techniques, and dimensional analysis to examine their dynamics. Thus we have
built a bridge between classic models in signal detection theory and nonlinear stochastic processes. Continuous stochastic models have been very useful in interpreting human  decision making in static environments \cite{bogacz06,gold07}. Dynamic environments offer a promising future direction for theory and experiments to probe the biophysical mechanisms that underlie decisions. \\
\vspace{-2mm}

\noindent
{\bf Acknowledgements.}  We thank Eric Shea--Brown, Shawn Gu, and the anonymous reviewers for helpful comments. Funding was provided by NSF-DMS-1311755 (ZPK); NSF-DMS-1517629 (KJ and ZPK); NSF/NIGMS-R01GM104974 (AV-C and KJ); and NSF-DMS-1122094 (KJ). \\[1ex]

\appendix

\begin{center} {\bf APPENDIX} \end{center}

\noindent
In this appendix, we present the derivations for the probability update formulas and their approximations discussed in the main text. We begin by deriving the update expression for the probability ratio, $R_n,$  in the case of two alternatives in a changing environment.  The result is a nonlinear recursive equation. Subsequently, we show how to approximate the log-likelihood ratio, $y_n = \ln R_n,$ using a SDE. To make the approximation precise, it is key to view the discrete equation for $y_n$  as a family of equations parameterized by the time interval, $\Delta t,$ over which each observation, $\xi_n,$ is made \cite{bogacz06}. Furthermore, we extend our derivations to multiple  ($N>2$) alternatives, and show that the
log probability updates can be approximated by a nonlinear system of SDEs in the continuum limit. With the appropriate scaling of the probabilities, $f_i(\xi) = {\rm Pr}(\xi|H_i),$  we can make precise the correspondence between the discrete and continuum models of posterior probability evolution. Lastly, we present a derivation for the stochastic integro--differential equation that represents the log probability for a continuum of possible environmental states, $\theta \in [a,b]$.

Note that throughout the appendix, we use notation involving a subscript $\Delta t$. This helps us define a family of stochastic processes indexed by the spacing between observations $\Delta t = t_n - t_{n-1}$. For instance, $f_{\Delta t , \pm}(\xi)$ represents the probability of an observation, $\xi,$ in environmental state  $H_{\pm}$  (or, in the language of statistics, when hypothesis $H_{\pm}$ holds).  This probability changes with the timestep $\Delta t$. This approach allows us to properly take the continuum limit $\Delta t \to 0$. However, for simplicity we refrain from using this notation in the main text.  Rather, we treat the limiting SDEs as approximations of discrete update processes. Also, we slightly abuse notation and
write $f_i(\xi) = {\rm Pr}(\xi|H_i),$ even when $\xi$ is a continuous random variable.

\section{Likelihood ratio for two alternatives}
\label{llr2app}
We begin by deriving the recursive update equation for the probabilities $L_{n,\pm}:={\rm Pr}(H(t_n)=H_{\pm}|\xi_{1:n})$
associated with each alternative $H_{\pm}$, where each observation (measurement), $\xi_i,$ is made at time $t_i$. This is the probability that alternative $H_{\pm}$ is true at time $t_n$, given that the series of observations $\xi_{1:n}$ has been made. Importantly, the underlying truth changes stochastically, and in a memoryless way, with transition probabilities given by $\epsilon_{\Delta t, \pm}:={\rm Pr}(H(t_n)=H_{\mp}|H(t_{n-1})=H_{\pm})$, so that  ${\rm Pr}(H(t_n)=H_{\pm}|H(t_{n-1})=H_{\pm}) = 1-\epsilon_{\Delta t, \pm}$. We begin by examining the probability $L_{n,+}$ associated with the alternative $H_+$. Using Bayes' rule and the law of total probability (Ch. 3 in \cite{rozanovbook}) we can relate the current probability, $L_{n,+},$ to the conditional probabilities at the time of the previous observation, $t_{n-1}$:
\begin{align*}
L_{n,+} &= \frac{1}{{\rm Pr}(\xi_{1:n})} \sum_{s={\pm}} {\rm Pr}(\xi_{1:n}|H(t_n)=H_+,H(t_{n-1})=H_s) \\ & \hspace{3cm} \times {\rm Pr}(H(t_n)=H_+,H(t_{n-1})=H_s),
\end{align*}
marginalizing over the joint distribution for the current $H(t_n)$ and previous $H(t_{n-1})$ environmental states. Next, we can apply the definition of the conditional probability ${\rm Pr}(H(t_n) = H_+|H(t_{n-1})=H_s)$ to write
\begin{align*}
L_{n,+} &= \frac{1}{{\rm Pr}(\xi_{1:n})} \sum_{s={\pm}} {\rm Pr}(\xi_{1:n} | H(t_n) = H_+, H(t_{n-1})=H_s) \\ & \hspace{2.5cm} \times {\rm Pr} (H(t_n) = H_+|H(t_{n-1})=H_s) {\rm Pr}(H(t_{n-1}) = H_s).
\end{align*}
Furthermore, we can split the joint condition on the first term by using the fact that the probability of making an observation $\xi_n$ is independent of $H(t_{n-1})$ when we condition on the present state $H(t_n) =H_+$ of the environment, so ${\rm Pr}(\xi_{1:n} | H(t_n) = H_+ , H(t_{n-1}) = H_s) = {\rm Pr}(\xi_n|H(t_n)=H_+){\rm Pr}(\xi_{1:n-1} | H (t_{n-1})=H_s)$ and
\begin{align*}
L_{n,+} &= \frac{1}{{\rm Pr}(\xi_{1:n})} \sum_{s = \pm} {\rm Pr}(\xi_n | H(t_n) = H_+) {\rm Pr}(\xi_{1:n-1}|H(t_{n-1}) = H_s) \\ & \hspace{2.5cm} \times {\rm Pr}(H(t_n) = H_+| H(t_{n-1}) = H_s) {\rm Pr}(H(t_{n-1}) = H_s).
\end{align*}
Lastly, we apply Bayes' rule to switch the order of ${\rm Pr}(\xi_{1:n-1}|H(t_{n-1})=H_s)$, yielding terms involving $L_{n-1,s} = {\rm Pr}(H(t_{n-1})=H_s|\xi_{1:n-1})$. In addition, we use ${\rm Pr}(H(t_n) = H_+|H(t_{n-1}) = H_+) = 1- \epsilon_{\Delta t, +}$ and ${\rm Pr}(H(t_n) = H_+|H(t_{n-1}) = H_-) = \epsilon_{\Delta t ,-}$ so that
\begin{align} \label{eqn:Lp}
L_{n,+} &= \frac{{\rm Pr}(\xi_{1:n-1}) {\rm Pr}(\xi_n | H(t_n) = H_+)}{{\rm Pr}(\xi_{1:n})} \left( \left( 1 - \epsilon_{\Delta t, +} \right) L_{n-1,+} + \epsilon_{\Delta t, -} L_{n-1,-} \right),
\end{align}
where $L_{0,+} = {\rm Pr}(H(t_0) = H_+)$.

Similarly, we can obtain an update equation for the probability $L_{n,-}$ of the alternative $H_-$ at time $t_n$: 
\begin{equation}\label{eqn:Lm}
L_{n,-} =\frac{{\rm Pr}(\xi_{1:n-1}){\rm Pr}(\xi_n|H(t_n)=H_-)}{{\rm Pr}(\xi_{1:n})}  \left( \epsilon_{\Delta t, +} L_{n-1,+} + (1-\epsilon_{\Delta t, -})L_{n-1,-} \right),
\end{equation}
where $L_{0,-}={\rm Pr}(H(t_0) = H_-)$.

From Eqs.~\eqref{eqn:Lp} and \eqref{eqn:Lm}, the  ratio $R_n = L_{n,+}/L_{n,-}$ is readily seen to satisfy the recursive equation
\begin{align}\label{eqn:llr_N2}
R_n& =\frac{f_{\Delta t, +}(\xi_n)}{f_{\Delta t, -}(\xi_n)}
\frac{(1-\epsilon_{\Delta t, +}) R_{n-1} + \epsilon_{\Delta t, -}}{\epsilon_{\Delta t, +} R_{n-1} + 1-\epsilon_{\Delta t, -}},
\end{align}
where $f_{\Delta t, \pm}(\xi_n)={\rm Pr}(\xi_n|H(t_n)=H_{\pm})$ is the distribution for each choice parameterized by the timestep $\Delta t = t_n - t_{n-1}$, and $R_0=\frac{{\rm Pr}(H_+,t_0)}{{\rm Pr}(H_-,t_0)}$.

\section{The continuum limit for the log-likelihood ratio of two alternatives}
\label{cont2app}
In this section, we derive a continuum equation for the log-likelihood ratio $y_n :=\ln R_n$. We will proceed by first defining a family of stochastic difference equations for $y_n$, which are parameterized by the timestep, $\Delta t = t_{n} - t_{n-1}$, between pairs of observations. By choosing an appropriate parameterization, we obtain a continuum limit that is a SDE. To begin, we divide both sides of Eq.~\eqref{eqn:llr_N2} by $R_{n-1}$
and take logarithms to yield
\begin{align} \label{eqn:dy_disc}
y_n-y_{n-1}& =\ln \frac{f_{\Delta t, +}(\xi_n)}{f_{\Delta t, -}(\xi_n)}+
\ln\frac{1-\epsilon_{\Delta t, +} + \epsilon_{\Delta t, -}e^{-y_{n-1}}}{1-\epsilon_{\Delta t, -} + \epsilon_{\Delta t, +} e^{y_{n-1}}}.
\end{align}
Following~\cite{bogacz06,bitzer14}, we assume that the time interval between individual observations, $\Delta t,$ 
is small. Denote by $\Delta y_{n} = y_{n} - y_{n-1}$ the change in the log-likelihood ratio due to the observation at time $t_{n}$. 
By assumption, the probability that the environment changes between two observations scales linearly with $\Delta t$ up to higher order terms, so that $\epsilon_{\Delta t, \pm} := \Delta t \epsilon_{\pm} + o(\Delta t)$.
Omitting higher order terms in $\Delta t$, Eq.~\eqref{eqn:dy_disc} can then be rewritten as
\begin{equation*}
\Delta y_{n}   = \ln \frac{f_{\Delta t , +}(\xi_{n})}{f_{\Delta t ,  -}(\xi_{n})} +  \ln(1+\Delta t (-\epsilon_+ +\epsilon_- e^{-y_{n-1}}))  -\ln(1+\Delta t(-\epsilon_- + \epsilon_+ e^{y_{n-1}})).
\end{equation*}
Since we assumed $\Delta t \ll 1$, we can use the approximation $\ln(1+a)\approx a$ which is valid to linear order in $|a|\ll 1$. We also assume that the change in the log-likelihood ratio,
$\Delta y_{n},$ is small over the time interval $\Delta t$, so $y_{n-1}$ can be replaced by $y_n$ on the right-hand side of the equation. We obtain
\begin{align}\label{eqn:dy_llr_N2}
\Delta y_{n} & \approx \ln \frac{f_{\Delta t ,  +}(\xi_n)}{f_{\Delta t , -}(\xi_n)} +
\Delta t ( \epsilon_- (e^{-y_{n}} + 1) -\epsilon_{+}(1 + e^{y_{n}})) \nonumber \\
& =
 {\rm E}_{\xi} \left[\left. \ln \frac{f_{\Delta t ,  +}(\xi_n)}{f_{\Delta t , -}(\xi_n)} \right| H(t_n) \right] +  \left(  \ln \frac{f_{\Delta t ,  +}(\xi_n)}{f_{\Delta t ,  -}(\xi_n)} - {\rm E}_{\xi} \left[\left.  \ln \frac{f_{\Delta t ,  +}(\xi_n)}{f_{\Delta t ,  -}(\xi_n)} \right| H(t_n) \right] \right) \nonumber \\ & \hspace{4.3cm} +  \Delta t ( \epsilon_- (e^{-y_{n}} + 1) -\epsilon_+(1 + e^{y_{n}})),
\end{align}
where we have conditioned on the state of the environment, $H(t_n)=H_\pm$ at time $t_n$. Replacing the index $n$, with the time $t,$ we can therefore write
\begin{equation}
\Delta y_t  \approx \Delta t g_{\Delta t}(t) +\sqrt{\Delta t}\rho_{\Delta t}(t) \eta +  \Delta t ( \epsilon_- (e^{-y_t} + 1) -\epsilon_+(1 + e^{y_t})),  \label{eq:dtp}
\end{equation}
where $\eta$ is random variable with standard normal distribution, and
\begin{align} \label{E:parameters}
& g_{\Delta t}(t)  := \frac{1}{\Delta t} \text{E}_{\xi} \left[ \left. \ln \frac{f_{\Delta t ,  +}(\xi)}{f_{\Delta t , -}(\xi)} \right| H(t) \right]  \nonumber\\
&  
\rho^2_{\Delta t}(t):= \frac{1}{\Delta t} \text{Var}_{\xi} \left[ \left. \ln \frac{f_{\Delta t , +}(\xi)}{f_{\Delta -}(\xi)} \right| H(t) \right].
\end{align}
As before, ${\rm E}_{\xi} \left[ F(\xi) \bigg| H(t) \right]$ is the expectation of $F(\xi)$ when $\xi$ is drawn from the distribution $f_{\pm}(\xi)$ associated with the current state $H(t) = H_{\pm}$. Clearly, the drift $g_{\Delta t}$ and variance $\rho_{\Delta t}^2$ will diverge or vanish unless $f_{\Delta t ,  \pm}(\xi)$ are scaled appropriately in the $\Delta t \to 0$ limit. We discuss different ways of introducing such a scaling in the next section. 

Assuming that we have well-defined limits $g(t) : = \lim_{\Delta t \to 0} g_{\Delta t} (t)$ and $\rho^2(t) : = \lim_{\Delta t \to 0} \rho_{\Delta t}^2 (t)$, the discrete-time stochastic process, Eq.~(\ref{eq:dtp}), approaches the SDE
\begin{align}\label{eqn:sde2}
\d y =  g(t)\d t +\rho(t) \d W_t + ( \epsilon_- (e^{-y} + 1) -\epsilon_+(1 + e^{y}))  \d t,
\end{align}
where $W_t$ is a standard Wiener process. This limit holds in the sense of distributions.
Roughly, the smaller $\Delta t$ is, the closer the distributions of the random variables $y_{n}$  and  $y(t_n)$ whose evolutions are described by Eq.~\eqref{eqn:dy_disc}, and Eq.~\eqref{eqn:sde2}, respectively. This correspondence
can be made precise using the Donsker Invariance Principle (p.520 in \cite{Billingsley_book}).

In sum, Eq.~\eqref{eqn:sde2}, can be viewed as an approximation of  the logarithm of the likelihood ratio whose evolution is given exactly by Eq.~\eqref{eqn:llr_N2}.  For a fixed interval $\Delta t$, the parameters of the two equations are related via 
Eq.~\eqref{E:parameters}, and $\epsilon_{\Delta t, \pm} / \Delta t= \epsilon_{\pm}$. 


\section{Precise correspondence}
\label{preciseapp}
We now discuss two approaches in which the  correspondence between Eqs.~\eqref{eqn:dy_disc} and \eqref{eqn:sde2} can be made exact. We choose a specific scaling for the drift and variance arising from each observation, $\xi_n$.  Suppose that over the time interval $\Delta t$, an observation, $\xi_n,$ is a result of
$r \Delta t$ separate observations -- for example the measurement of the direction of $r \Delta t$ different moving dots~\cite{gold07}. 
In this case the estimate of the average of the individual measurements -- \emph{e.g.}, the average of the
velocities of dots in a display -- will have both a mean and a variance that increase linearly with $\Delta t$. 

As a concrete example we can compute $g(t)$ and $\rho (t)$ in SDE~(\ref{eqn:sde2}) when 
observations, $\xi_n$, follow normal distributions with mean and variance scaled by $\Delta t$,
\begin{align*}
f_{\Delta t,\pm} (\xi) = \frac{1}{\sqrt{2 \pi \Delta t \sigma^2}} \e^{-(\xi - \Delta t\mu_{\pm})^2/(2 \Delta t \sigma^2)}.
\end{align*}
Using Eq.~\eqref{E:parameters} it is then straightforward to compute \cite{bogacz06,bitzer14},
\begin{align*}
g_{\Delta t } (t) & 
=  \pm  \frac{(\mu_+ - \mu_-)^2}{2 \sigma^2} = g_{\pm}, \\
\rho_{\Delta t}^2 (t) &
=    \frac{(\mu_+ - \mu_-)^2}{\sigma^2} = \rho^2,
\end{align*}
and note that $g(t) \in \left\{ g_+, g_- \right\}$ is a telegraph process  (\emph{e.g.}, p.~77~in~\cite{gardiner04}) with the probability masses $P(g_{\pm},t)$ evolving according to the master equation $P_t (g_{\pm},t) = \mp \epsilon_+ P(g_+,t) \pm \epsilon_- P(g_-,t)$.  In this case $\rho^2(t) = \rho^2$ remains constant. 

 

More generally, we can obtain an identical result by considering that each observation made on a time interval consists of a number of sub-observations, each with statistics that scale with the length of the interval and the number of sub-observations. We define a family of stochastic processes parameterized by $k$, the number of sub-observations made in an interval of length $\Delta t$. 
As above, when $k=1$, we assume that an observation $\xi_n$ is the result of $r \Delta t$ separate observations. Assuming $r$ is large, note that for $k>1$  each of the $k$ subobservations contain roughly $r_k = \lfloor{r\Delta t/k} \rfloor$ observations with mean and variance that scale linearly with $r_k\propto \Delta t/k$. 
We can achieve this by approximating $\ln \frac{f_{\Delta t,+}(\xi_n)}{f_{\Delta t,-}(\xi_n)}$ in Eq.~\eqref{eqn:dy_llr_N2} by the family of stochastic processes parameterized by $k$ \cite{bogacz06}
\begin{equation*}
 \sum_{l=1}^k \frac{\Delta t}{k}\ln \frac{f_+(\xi_l)}{f_-(\xi_l)} +  \sum_{l=1}^k \frac{\sqrt{\Delta t}}{\sqrt{k}}   \left(\ln \frac{f_+(\xi_l)}{f_-(\xi_l)} - {\rm E}_{\xi}\left[\left.\ln \frac{f_+(\xi)}{f_-(\xi)} \right| H(t)\right]  \right).
\end{equation*}
The scaling in this approximation guarantees that the drift is given by the limit
$$g(t)=\lim_{\Delta t \rightarrow 0} g_{\Delta t}(t)=\lim_{\Delta t\rightarrow 0} \frac{1}{\Delta t} {\rm E}_{\xi} \left[ \left. \ln \frac{f_{\Delta t ,  +}(\xi)}{f_{\Delta t ,  -}(\xi)} \right| H(t) \right] = {\rm E}_{\xi} \left[\left. \ln \frac{f_+(\xi)}{f_-(\xi)}\right|H(t) \right]$$ 
and the variance
$$\rho^2(t)=\lim_{\Delta t\rightarrow 0}\rho^2_{\Delta t}=\lim_{\Delta t\rightarrow 0}\frac{1}{\Delta t} {\rm Var}_{\xi} \left[ \left. \ln \frac{f_{\Delta t ,  +}(\xi)}{f_{\Delta t ,  -}(\xi)} \right| H(t) \right] =  {\rm Var}_{\xi} \left[\left. \ln \frac{f_+(\xi)}{f_-(\xi)}\right|H(t) \right].$$ Furthermore, as $k\rightarrow \infty$, by the Central Limit Theorem, 
\begin{align*}
\Delta y_t \approx &  \sum_{l=1}^k \frac{\Delta t}{k}\ln \frac{f_+(\xi_l)}{f_-(\xi_l)} + \sum_{l=1}^k \frac{\sqrt{\Delta t}}{\sqrt{k}}  \left(\ln \frac{f_+(\xi_l)}{f_-(\xi_l)} - {\rm E}_{\xi}\left[\left. \ln \frac{f_+(\xi)}{f_-(\xi)} \right| H(t)\right]  \right) \\ & \hspace{2.6cm} +   \Delta t ( \epsilon_- (e^{-y_t} + 1) -\epsilon_+(1 + e^{y_t}))
\end{align*}
converges in distribution to
\begin{equation*}
\Delta y_t  \approx \Delta t g(t) +\sqrt{\Delta t}\rho(t) \eta + \Delta t ( \epsilon_- (e^{-y_t} + 1) -\epsilon_+(1 + e^{y_t})),
\end{equation*}
where $\eta$ is a standard normal random variable. Taking the limit $\Delta t\rightarrow 0$  yields Eq.~\eqref{eqn:sde2}. When observations follow Gaussian distributions, $f_\pm\sim \mathcal{N}(\pm \mu,\sigma^2)$, then $g(t)=\pm 2\mu^2 / \sigma^2$, $\rho=2 \mu / \sigma$, and 
\begin{align*}
\d y =  \left[ g(t) + ( \epsilon_- (e^{-y} + 1) -\epsilon_+(1 + e^{y})) \right]  \d t +\rho \, \d W,
\end{align*}
where $dW$ is a standard white noise process.

\section{Continuum limit for log probabilities with multiple alternatives}
\label{llmultiapp}
We now describe the calculation of the continuum limit of the recursive system defining the evolution of the probabilities $L_{n,i} = {\rm Pr} (H(t_n)=H_i | \xi_{1:n})$ 
of one among multiple alternatives (environmental states), $H_i,$ $i=1,..,N$. The state of the environment, and equivalently the correct choice at time $t$, again change stochastically. We assume that the transitions between the alternatives are memoryless, with transition rates $\epsilon_{\Delta t, ij} : = {\rm Pr}(H(t_n)=H_i | H(t_{n-1})=H_j)$. Using Bayes' rule and rearranging terms (analogous to the derivation of Eqs.~\eqref{eqn:Lp} and \eqref{eqn:Lm}), we can express each probability $L_{n,i}$ in terms the  probability at the time of the previous observation, $L_{n-1,j},$
\begin{align*}
L_{n,i} = \frac{{\rm Pr}(\xi_{1:n-1})}{{\rm Pr} ( \xi_{1:n} )} {\rm Pr}(\xi_n|H_i,t_n) \sum_{j=1}^N \epsilon_{\Delta t, ij} L_{n-1,j}.
\end{align*}
Since we are only interested in comparing the magnitude of the probabilities, we can drop the common prefactor $\frac{{\rm Pr}(\xi_{1:n-1})}{{\rm Pr} ( \xi_{1:n} )},$ and use the fact that $\sum_{j=1}^N \epsilon_{\Delta t,ji} = 1$ (since $\epsilon_{\Delta t, ij}$ is a left stochastic matrix) to write $\epsilon_{\Delta t, ii} = 1 - \sum_{j \neq i} \epsilon_{\Delta t,ji} $ and obtain
\begin{equation}
L_{n,i}  = f_{\Delta t, i}(\xi_n)  \left( \left[ 1 - \sum_{j \neq i} \epsilon_{\Delta t, ji} \right] L_{n-1,i} +   \sum_{j \neq i} \epsilon_{\Delta t, ij} L_{n-1,j} \right), \label{multilike}
\end{equation}
where $f_{\Delta t, i}(\xi_n)={\rm Pr}(\xi_n|H_i,t_n)$. 
From Eq.~(\ref{multilike}), it follows that log of the rescaled 
probabilities, 
 $x_i:=\ln L_i$, satisfies the recursive relation
\begin{equation*}
x_{n,i}-x_{n-1,i}  = \ln f_{\Delta t, i}(\xi_n) +  \ln \left(1-\sum_{j\neq i} \epsilon_{\Delta t, ji}+
  \sum _{j\neq i}\epsilon_{\Delta t, ij}e^{x_{n-1,j}-x_{n-1,i}} \right).
\end{equation*}

To derive an approximating SDE, we denote by $\Delta x_{n, i} = x_{n,i} - x_{n-1,i}$, the change in the log probability due to an observation at time $t_n$. As before, we assume $\epsilon_{\Delta t, ij} := \Delta t \epsilon_{ij} + o(\Delta t)$ for $i\neq j$,  and drop the higher order terms, giving
\begin{equation*}
\Delta x_{n,i}  = \ln f_{\Delta t ,  i} (\xi_n) +   \ln \left(1-\sum_{j\neq i} \Delta t \epsilon_{ji}+
  \sum _{j\neq i} \Delta t \epsilon_{ij}e^{x_{n-1,j}-x_{n-1,i}} \right).
\end{equation*}
Assuming $\Delta t \ll 1$, we again use the approximation $\ln (1+a) \approx a$ for $|a| \ll 1$. We also assume that the change in the log probability, $|\Delta x_{n,i}| \ll 1$, is small over the time interval $\Delta t$, so that
\begin{align}\label{eqn:dyn_N}
\Delta x_{n,i} \approx & \ln f_{\Delta t ,  i} (\xi_n)  + \Delta t \sum_{j \neq i} \left( \epsilon_{ij} \e^{x_{n,j} - x_{n,i}} - \epsilon_{ji} \right) \nonumber \\
=& {\rm E}_{\xi} \left[ \ln f_{\Delta t ,  i}(\xi) | H(t_n) \right] +   \left( \ln f_{\Delta t ,  i}(\xi_n) - {\rm E}_{\xi} \left[ \ln f_{\Delta t ,  i}(\xi) | H(t_n)\right]  \right) \nonumber \\ & \hspace{3.15cm} +  \Delta t \sum_{j \neq i} \left( \epsilon_{ij} \e^{x_{n,j} - x_{n,i}} - \epsilon_{ji} \right),
\end{align}
where we condition on the current state of the environment $H(t_n) \in \left\{H_1, ..., H_N \right\}$.

Replacing the index $n$, by the time $t$, we can therefore write
\begin{equation}\label{eqn:dtp_N}
\Delta x_{t,i}  \approx  \Delta t g_{\Delta t,i}(t) + 
\sqrt{\Delta t} \rho_{\Delta t,i}(t)\eta_i + 
\Delta t \sum_{j \neq i} \left( \epsilon_{ij} \e^{x_{t,j} - x_{t,i}} - \epsilon_{ji} \right), 
\end{equation}
where $\eta_i$'s are correlated random variables with standard normal distribution
\begin{align*}
& g_{\Delta t,i}(t):= \frac{1}{\Delta t} \text{E}_{\xi} \left[ \left. \ln f_{\Delta t ,  i}(\xi)\right| H(t) \right] \\
& \rho^2_{\Delta t,i}(t):= \frac{1}{\Delta t} \text{Var}_{\xi} \left[ \left. \ln f_{\Delta t , i}(\xi)\right| H(t) \right].
\end{align*}
The correlation of $\eta_i$'s is given by 
\begin{equation*} 
{\rm Corr}_{\xi}[\eta_i,\eta_j]:= \text{Corr}_{\xi} \left[\left. \ln f_{\Delta t , i}(\xi),\ln f_{\Delta t , j}(\xi) \right| H(t) \right].
\end{equation*}
Note that Eq.~\eqref{eqn:dtp_N} is the multiple-alternative version of Eq.~\eqref{eq:dtp}. Equivalently, we can write Eq.~\eqref{eqn:dtp_N} as
\begin{equation*}
\Delta x_{t,i}  \approx   \Delta t g_{\Delta t,i}(t) + 
\sqrt{\Delta t} \widehat{W}_{\Delta t,i} +   \Delta t \sum_{j \neq i} \left( \epsilon_{ij} \e^{x_{t,j} - x_{t,i}} - \epsilon_{ji} \right),
\end{equation*}
where $\widehat{W}_{\Delta t}:= (\widehat{W}_{\Delta t, 1},\ldots,\widehat{W}_{\Delta t,N})$ follows a multivariate Gaussian distribution with mean zero and covariance matrix $\Sigma_{\Delta t}$ given by
\begin{equation*} 
\Sigma_{\Delta t, ij}= \frac{1}{\Delta t}\text{Cov}_\xi \left[\left. \ln f_{\Delta t , i}(\xi),\ln f_{\Delta t , j}(\xi) \right| H(t) \right].
\end{equation*}

Finally, taking the limit $\Delta t \to 0$, and assuming that the limits 
\begin{equation} \label{E:limits}
g_i(t) : = \lim_{\Delta t \to 0} g_{\Delta t ,  i} (t), \quad \text{and} \quad \Sigma_{ij}(t) : = \lim_{\Delta t \to 0} \Sigma_{\Delta t ,  ij} (t),
\end{equation} 
are well defined, we obtain the system of SDEs
\begin{align}
\d x_i = g_i(t) \d t + \d \widehat{W}_i (t)  + \sum_{j \neq i} \left( \epsilon_{ij} \e^{x_j - x_i} - \epsilon_{ji} \right) \d t  ,  \label{multisde}
\end{align}
or equivalently as the vector system
\begin{align*}
\d \mathbf{x} = \mathbf{g} (t) \d t + \Lambda(t) \d \mathbf{W}_t  + K(\mathbf{x} ) \d t ,
\end{align*}
where $\mathbf{g}(t) = (g_1(t), ..., g_N(t))^T$ and $\Lambda(t) \Lambda(t)^T = \Sigma(t)$
are defined using the limits in Eq.~\eqref{E:limits}, $K_i(\mathbf{x}) = \sum_{j \neq i} \left( \epsilon_{ij} \e^{x_j - x_i} - \epsilon_{ji} \right)$,  and the components of $\mathbf{W}_t$ are independent Wiener processes. 
We can recover Eq.~(\ref{eqn:sde2}) by taking $N=2$, letting $y = x_1 - x_2$, and exchanging the indices $1$ and $2$ with $+$ and $-,$ respectively. 


As in the case of two alternatives, Eq.~\eqref{multisde} can be viewed as an approximation of  the logarithm of the probability 
 whose evolution is given exactly by Eq.~\eqref{multilike}.  For a fixed interval $\Delta t$, the parameters of these equations are related via 
Eq.~\eqref{multisde}, and $\epsilon_{\Delta t, ij} / \Delta t= \epsilon_{ij}$.

The limits $g_i(t) : = \lim_{\Delta t \to 0} g_{\Delta t ,  i} (t)$ and $\Sigma_{ij}(t) : = \lim_{\Delta t \to 0} \Sigma_{\Delta t ,  ij} (t)$ are defined when the statistics of the observations scale with $\Delta t$. As we argued above, this can be obtained by considering observations drawn from a normal distribution with mean and variance scaled by $\Delta t$:
\begin{align*}
f_{\Delta t,i} (\xi) = \frac{1}{\sqrt{2 \pi \Delta t \sigma^2}} \e^{-(\xi - \Delta t\mu_{i})^2/(2 \Delta t \sigma^2)}.
\end{align*}
Alternatively, the required scaling can also be obtained when each observation made on a time interval consists of a number of sub-observations, $(\xi_1,\ldots,\xi_k)$, with mean and variance scaled by $\frac{\Delta t}{k}$. To do so we approximate $\ln f_{\Delta t,i}(\xi_n)$ in Eq.~\eqref{eqn:dyn_N} by
\begin{align*}
 \sum_{l=1}^k \frac{\Delta t}{k}\ln f_i(\xi_l) + 
 \sum_{l=1}^k \frac{\sqrt{\Delta t}}{\sqrt{k}}\left(\ln f_i(\xi_l)- {\rm E}_{\xi}\left[\ln f_i(\xi)| H(t)\right]  \right).
\end{align*}

\section{Log-likelihood ratio for multiple alternatives}
\label{llrmultiapp}
We can also derive a continuum limit for the log-likelihood ratio for any two choices $i,j \in \left\{ 1, 2, ..., N \right\}$. From Eq.~(\ref{multilike}), the likelihood ratio $R_{n,ij} = L_{n,i}/L_{n,j}$.  We note that this will provide us with a matrix of stochastic processes.  We start with the recursive equation
\begin{equation*} 
R_{n,ij} =  \frac{f_{\Delta t, i} (\xi_n)}{f_{\Delta t, j}(\xi_n)} \frac{\left(1- \sum_{k \neq i} \epsilon_{\Delta t, ki} \right) R_{n-1,ij} + \sum_{k \neq i} \epsilon_{\Delta t, ik}R_{n-1,kj}}
{1 - \sum_{k \neq j} \epsilon_{\Delta t, kj} + \sum_{k \neq j} \epsilon_{\Delta t,jk}  R_{n-1,kj} }.  
\end{equation*}
We can thus derive the continuum equation for the log-likelihood ratio $y_{n, ij} : = \ln R_{n, ij}$, as we did in the case of two alternatives. 
Since $y_{ij}(t)$ is the difference 
 $y_{ij}(t)=x_{i}(t)-x_{j}(t)$, from Eq.~(\ref{multisde}) we obtain
\begin{equation*}
\d y_{ij} = (g_i(t)-g_j(t)) \d t + \d \widehat{W}_i (t)-\d \widehat{W}_j(t) + \sum_{k \neq i} \left( \epsilon_{ik} \e^{y_{ki}} - \epsilon_{ki} \right) \d t 
- \sum_{k \neq j} \left( \epsilon_{jk} \e^{y_{kj}} - \epsilon_{kj} \right) \d t
,
\end{equation*}
or 
 \begin{equation} \label{E:loglike_ratio}
\d y_{ij} =  \left[ g_{ij}(t) +  \left ( \sum_{k \neq j} \epsilon_{kj} -\sum_{k \neq i} \epsilon_{ki} + 
\sum_{k \neq i} \epsilon_{ ik}e^{y_{ki}}
 -  \sum_{k \neq j} \epsilon_{jk}  e^{y_{kj}}
 \right) \right] \d t  +  \d \widehat{W}_{ij} ,
\end{equation}
where 
$g_{ij}(t)=\text{E}_{\xi}\left[\left. \ln \frac{f_i(\xi)}{f_j(\xi)} \right| H(t) \right]$ and $\widehat{W}$ is a Wiener process with covariance matrix given by 
$\text{Cov}_\xi\left[\left.\widehat{W}_{ij},\widehat{W}_{i'j'}\right| H(t) \right]
=\text{Cov}_\xi \left[\left. \ln \frac{f_i(\xi)}{f_j(\xi)}, \ln \frac{f_{i'}(\xi)}{f_{j'}(\xi)}\right| H(t) \right]$.
We can also write Eq.~\eqref{E:loglike_ratio} in vector form
 \begin{align*} 
\d \mathbf{y} =  \mathbf{g} \d t  +  \Lambda(t)\d \mathbf{W}_t +
\mathbf{K}(\mathbf{y} ) \d t   ,
\end{align*}
where $\mathbf{K}_{ij}(\mathbf{y})= \sum_{k \neq j} \epsilon_{kj} -\sum_{k \neq i} \epsilon_{ki} + 
\sum_{k \neq i} \epsilon_{ ik}e^{y_{ki}} -  \sum_{k \neq j} \epsilon_{jk}  e^{y_{kj}}$, $\Lambda(t)\Lambda(t)^T=\Sigma(t)$ is the covariance matrix, and the components of $\mathbf{W}_t $ are independent Wiener processes.

\section{Log probabilities for a continuum of alternatives}
\label{contapp}
Finally, we examine the case where an observer must choose between a continuum of hypotheses $H_{\theta}$ where $\theta \in [a,b]$. Thus, we will first derive a discrete recursive equation for the evolution of the probabilities $L_{n, \theta} = {\rm Pr} (H(t_n)=H_\theta| \xi_{1:n})$. The state of the environment, the correct choice at time $t$, again changes according to a continuous time Markov process. We define this stochastically switching process through its transition rate function $\epsilon_{\Delta t, \theta \theta'}$, which is given for $\theta' \neq \theta$ as 
\begin{align*}
\int_{\theta_1}^{\theta_2} \epsilon_{\Delta t, \theta \theta'}  \d \theta  : = {\rm Pr}  \left( H(t_n) \in H_{[\theta_1, \theta_2 ]} \right| \left. H(t_{n-1}) = H_{\theta'} \right),
 \end{align*}
 where $H_{[\theta_1, \theta_2]}$ is the set of all states $H_{\theta}$ with $\theta$ in the interval $[\theta_1, \theta_2 ]$.
Thus, $\epsilon_{\Delta t, \theta \theta'}$ describes the probability of a transition over a timestep, $\Delta t$, from state $H_{\theta'}$ to some state $H_{\theta}$, with $\theta \in [\theta_1, \theta_2 ]$. This means that ${\rm Pr} (H(t_n)=H_{\theta} | H(t_{n-1})=H_{\theta})=1 - \int_a^b \epsilon_{\Delta t, \theta' \theta} \d \theta'$.
As in the derivation of the multiple alternative $2 \leq N < \infty$ case, we can express each probability $L_{n, \theta}$ at time $t_n$ in terms of the probabilities $L_{n-1,\theta'}$ at time $t_{n-1}$, so
\begin{align*}
L_{n, \theta} =& \frac{{\rm Pr}(\xi_{1:n-1})}{{\rm Pr}(\xi_{1:n})} 
{\rm Pr}(\xi_n|H(t_n)=H_{\theta}) \\
& \hspace{1cm} \times \left(\text{Pr}(H(t_n)=H_\theta|H(t_{n-1})=H_\theta)L_{n-1,\theta} + \int_a^b \epsilon_{\Delta t, \theta \theta'}
 L_{n-1,\theta'} \d \theta'
 \right).
\end{align*}
Notice that the sum from the $N<\infty$ case, as in Eq.~(\ref{multilike}), has been replaced with an integral over all possible hypotheses $H_{\theta'}$, $\theta' \in [a,b]$ and a term corresponding to the probability of the environment not changing. Again we drop the common factor $\frac{{\rm Pr}(\xi_{1:n-1})}{{\rm Pr}(\xi_{1:n})}$, since we wish to compare the magnitudes of the probabilities. 
We obtain
\begin{align}
L_{n, \theta} & = f_{\Delta t, \theta}(\xi_n) 
\left( \left[ 1 - \int_a^b \epsilon_{\Delta t, \theta' \theta} \d \theta' \right] L_{n-1,\theta} +   \int_a^b \epsilon_{\Delta t, \theta \theta'} L_{n-1,\theta'} \d \theta' \right), \label{contlike}
\end{align}
where $f_{\Delta t, \theta}(\xi_n)={\rm Pr}(\xi_n|H(t_n)=H_{\theta} )$. From Eq.~(\ref{contlike}), we can thus derive a recursive relation for the log of the rescaled probabilities $x_{n,\theta} : = \ln L_{n, \theta}$ in terms of $x_{n-1,\theta}$ so
\begin{align*}
x_{n, \theta} - x_{n-1,\theta} = \ln f_{\Delta t, \theta} (\xi_n) + \ln \left( 1 -  \int_a^b \epsilon_{\Delta t, \theta' \theta} \d \theta' +   \int_a^b \epsilon_{\Delta t, \theta \theta'} \e^{x_{n-1,\theta'} - x_{n-1,\theta}} \d \theta' \right).
\end{align*}
To approximate this discrete-time stochastic process with a SDE, we denote by $\Delta x_{n, \theta} = x_{n,\theta} - x_{n-1,\theta}$, the change in  log probability due to the observation at time $t_n$. Furthermore, we assume $\epsilon_{\Delta t, \theta \theta'} : = \Delta t \epsilon_{\theta \theta'} + o (\Delta t)$ and drop higher order terms,
\begin{align*}
\Delta x_{n, \theta} & = \ln f_{\Delta t, \theta} (\xi_n) + \ln \left( 1 -  \int_a^b \Delta t \epsilon_{\theta' \theta} \d \theta'  +  \int_a^b \Delta t \epsilon_{\theta \theta'} \e^{x_{n-1,\theta'} - x_{n-1,\theta}} \d \theta' \right).
\end{align*}
Assuming $\Delta t \ll 1$, we again use the approximation $\ln (1+a) \approx a$ for $|a| \ll 1$. Assuming $|\Delta x_{n, \theta}| \ll 1$, 
\begin{align*}
\Delta x_{n, \theta} & \approx \ln f_{\Delta t, \theta} (\xi_n) + \Delta t \int_a^b \left( \epsilon_{\theta \theta'} \e^{x_{n,\theta'} - x_{n, \theta}} - \epsilon_{\theta' \theta} \right) \d \theta' \\
&= {\rm E}_{\xi} \left[ f_{\Delta t, \theta} (\xi)| H(t_n) \right] +  \left( \ln f_{\Delta t, \theta}(\xi_n) - {\rm E}_{\xi} \left[ \ln f_{\Delta t, \theta} (\xi) | H(t_n)\right] \right) \\ & \hspace{3.33cm} +   \Delta t \int_a^b \left( \epsilon_{\theta \theta'} \e^{x_{n,\theta'} - x_{n, \theta}} - \epsilon_{\theta' \theta} \right) \d \theta',
\end{align*}
conditioned on the current state of the environment $H (t_n) = H_{\varphi}$ where $\varphi \in [a,b]$. 

Exchanging the index $n$ with the time, $t$, we can therefore write
\begin{align}\label{eqn:dtp_theta}
\Delta x_{t,\theta} \approx & \Delta t g_{\Delta t,\theta}(t) + 
\sqrt{\Delta t} \rho_{\Delta t,\theta}(t)\eta_\theta + \Delta t \int_a^b \left( \epsilon_{\theta \theta'} \e^{x_{t,\theta'} - x_{t, \theta}} - \epsilon_{\theta' \theta} \right) \d \theta'
, 
\end{align}
where $\eta_\theta$'s are correlated random variables which marginally follow a standard normal distribution, and
\begin{align*}
& g_{\Delta t,\theta}(t):= \frac{1}{\Delta t} \text{E}_{\xi} \left[ \left. \ln f_{\Delta t ,  \theta}(\xi)\right| H(t) \right], \\
& \rho^2_{\Delta t,\theta}(t):= \frac{1}{\Delta t} \text{Var}_{\xi} \left[ \left. \ln f_{\Delta t , \theta}(\xi)\right| H(t) \right].
\end{align*}
The correlation of $\eta_i$'s is given by 
\begin{equation*} 
{\rm Corr}_{\xi}[\eta_\theta,\eta_{\theta'}]:= \text{Corr}_{\xi} \left[\left. \ln f_{\Delta t , \theta}(\xi),\ln f_{\Delta t , \theta'}(\xi) \right| H(t) \right].
\end{equation*}
 Equivalently, we can write Eq.~\eqref{eqn:dtp_theta} as
\begin{align*}
\Delta x_{t,\theta} & \approx  \Delta t g_{\Delta t,\theta}(t) + 
\sqrt{\Delta t} \widehat{W}_{\Delta t,\theta} + \Delta t \int_a^b \left( \epsilon_{\theta \theta'} \e^{x_{t,\theta'} - x_{t, \theta}} - \epsilon_{\theta' \theta} \right) \d \theta'
,
\end{align*}
where $\widehat{W}_{\Delta t}:= (\widehat{W}_{\Delta t, \theta})_{\theta\in[a,b]}$. For $\theta \in [a,b]$, $\widehat{W}_{\Delta t, \theta}$ is a Gaussian process in the sense that any finite subset of points $\{ \theta_1, ..., \theta_n \} \in [a,b]$ have a multivariate Gaussian distribution with mean zero and covariance, $\Sigma_{\Delta t, \theta \thetaÕ},$ given by
\begin{equation*}
\Sigma_{\Delta t, \theta\theta'}= \frac{1}{\Delta t}\text{Cov}_\xi \left[\left. \ln f_{\Delta t , \theta}(\xi),\ln f_{\Delta t , \theta'}(\xi) \right| H(t) \right].
\end{equation*}

Finally, taking the limit $\Delta t \to 0$, and assuming that the limits 
\begin{equation} \label{E:limits_theta}
g_\theta(t) : = \lim_{\Delta t \to 0} g_{\Delta t ,  \theta} (t), 
\hspace*{.1in} \text{and} \hspace*{.1in}  \Sigma_{\theta\theta'}(t) : = \lim_{\Delta t \to 0} \Sigma_{\Delta t ,  \theta\theta'} (t),
\end{equation} 
are well defined, we obtain the system of SDEs
\begin{align}
\d x_\theta = g_\theta(t) \d t + \d \widehat{W}_\theta (t)  + \int_a^b \left( \epsilon_{\theta \theta'} \e^{x_{\theta'} - x_{\theta}} - \epsilon_{\theta' \theta} \right) \d \theta' \d t , \label{app:side}
\end{align}
or equivalently as the system of SDEs
\begin{align*}
\d \mathbf{x} = \mathbf{g} (t) \d t + \Lambda(t) \d \mathbf{W}_t + K(\mathbf{x} ) \d t   ,
\end{align*}
where $\mathbf{g}(t) = \big(g_\theta(t)\big)_{\theta\in[a,b]}$ and  $\Lambda(t) \Lambda(t)^T = \Sigma(t)$
are defined using the limits in Eq.~\eqref{E:limits_theta}, $K(\mathbf{x}) = \int_a^b \left( \epsilon_{\theta \theta'} \e^{x_{\theta'} - x_{ \theta}} - \epsilon_{\theta' \theta} \right) \d \theta'$,  and the components of $\mathbf{W}_t$ are independent Wiener processes. 

While we have formally taken the limit of the discrete Eq.~(\ref{eqn:dtp_theta}), it is important to note that establishing the well-posedness of stochastic integrodifferential equations is not straight-forward. Conditions for the existence and uniqueness of solutions to certain nonlinear stochastic partial differential equations (SPDEs) are demonstrated in Ch.7 of \cite{dapratobook}. This approach considers the solutions to SPDEs to be random processes that take their values in a Hilbert space of functions. Recently, this concept has been extended to provide general conditions on the constituent functions of stochastic neural fields to ensure the existence of solutions \cite{kuehn14,faugeras14}. The form of stochastic neural fields is closely related to Eq.~(\ref{app:side}), since both types of equation possess a linear drift and a convolution defining a nonlocal coupling between their state variables. It may be possible to utilize these previous approaches to establish the existence and uniqueness of solutions to Eq.~(\ref{app:side}) in future studies.

\bibliographystyle{siam}
\bibliography{leaky}

\end{document}